\documentclass[twocolumn,tighten,twocolappendix]{aastex63}
\pdfoutput=1

\usepackage{color, hyperref}
\usepackage[normalem]{ulem}
\usepackage{amsmath}
\usepackage{fontawesome}

\newcommand{\code}[1]{\texttt{#1}}

\newcommand{\transpose}[1]{{#1}^{\!\mathsf T}}
\newcommand{\given}{\,|\,}
\renewcommand{\det}[1]{||{#1}||}
\newcommand{\conv}{\,\otimes\,}
\renewcommand{\det}[1]{||{#1}||}
\newcommand{\like}{\mathcal{L}}
\newcommand{\norm}{\mathcal{N}}

\newcommand{\fout}{{\rm f}_{\rm out}}
\newcommand{\unit}[1]{\mathrm{#1}}
\newcommand{\feh}{\log (\mathrm{Z}/\mathrm{Z}_\odot)}

\newcommand{\msun}{M_\odot}
\newcommand{\km}{\unit{km}}
\newcommand{\s}{\unit{s}}
\newcommand{\kms}{\km\,\s^{-1}}
\newcommand{\tage}{t_{age}}
\newcommand{\ba}{\boldsymbol{\alpha}}
\newcommand{\bdelt}{\boldsymbol{\Delta}}
\newcommand{\bg}{\boldsymbol{G}}

\newcommand{\pro}{\textsc{Prospector}}
\newcommand{\emcee}{\texttt{emcee}}
\newcommand{\fsps}{\textsc{FSPS}}
\newcommand{\pyfs}{\textsc{python-FSPS}}
\newcommand{\sedpy}{\texttt{sedpy}}

\newcommand{\dynesty}{\texttt{dynesty}}

\newcommand{\gnz}{GNz-11}

\usepackage{graphicx}
\DeclareGraphicsExtensions{.png,.pdf}
\graphicspath{{./}{figures/}}
\usepackage[]{animate}

\shorttitle{Prospector}
\shortauthors{Johnson et al.}
\begin{document}

\title{Stellar Population Inference with Prospector}

\author[0000-0002-9280-7594]{Benjamin~D.~Johnson}
\affiliation{Center for Astrophysics $|$ Harvard \& Smithsonian, 60 Garden Street, Cambridge, MA 02138, USA}
\author[0000-0001-6755-1315]{Joel~Leja}
\affiliation{Department of Astronomy \& Astrophysics, The Pennsylvania
State University, University Park, PA 16802, USA}
\affiliation{Institute for Computational \& Data Sciences, The Pennsylvania State University, University Park, PA, USA}
\affiliation{Institute for Gravitation and the Cosmos, The Pennsylvania State University, University Park, PA 16802, USA}
\author[0000-0002-1590-8551]{Charlie~Conroy}
\affiliation{Center for Astrophysics $|$ Harvard \& Smithsonian, 60 Garden Street, Cambridge, MA 02138, USA}
\author[0000-0003-2573-9832]{Joshua~S.~Speagle}
\altaffiliation{Banting Fellow}
\affiliation{Department of Statistical Sciences, University of Toronto, Toronto, ON M5S 3G3, Canada}
\affiliation{David A. Dunlap Department of Astronomy \& Astrophysics, University of Toronto, Toronto, ON M5S 3H4, Canada}
\affiliation{Dunlap Institute for Astronomy \& Astrophysics, University of Toronto, Toronto, ON M5S 3H4, Canada}
\affiliation{Center for Astrophysics $|$ Harvard \& Smithsonian, 60 Garden Street, Cambridge, MA 02138, USA}


\begin{abstract}
Inference of the physical properties of stellar populations from observed photometry and spectroscopy is a key goal in the study of galaxy evolution.
In recent years the quality and quantity of the available data has increased, and there have been corresponding efforts to increase the realism of the stellar population models used to interpret these observations.
Describing the observed galaxy spectral energy distributions in detail now requires physical models with a large number of highly correlated parameters.
These models do not fit easily on grids and necessitate a full exploration of the available parameter space.
We present \pro{}, a flexible code for inferring stellar population parameters from photometry and spectroscopy spanning UV through IR wavelengths.
This code is based on forward modeling the data and Monte Carlo sampling the posterior parameter distribution, enabling complex models and exploration of moderate dimensional parameter spaces.
We describe the key ingredients of the code and discuss the general philosophy driving the design of these ingredients.
We demonstrate some capabilities of the code on several datasets, including mock and real data.
\textbf{  \href{https://github.com/bd-j/exspect}{\faGithub}}
\end{abstract}


\section{Introduction}
Fitting galaxy spectral energy distributions (SEDs) is the primary method for converting galaxy observations into inferred physical properties. These physical properties can in turn be compared to predictions from theories of galaxy evolution. Accordingly, galaxy SED fitting has a rich history as one of the primary methods of hypothesis testing in extragalactic astronomy (see reviews by \citealt{faber77,tinsley80,walcher11,conroy13}).

There are three key components of SED fitting: a physical model, a set of observations such as photometry or spectroscopy, and a statistical inference framework which connects the model to the observations. The physical models are typically chosen to be stellar population synthesis codes, which combine inputs such as isochrones, stellar spectra, an initial mass function (IMF), a star formation and chemical evolution history, dust attenuation, and (sometimes) dust and nebular emission to create the SEDs of complex stellar populations. There are many publicly available stellar population synthesis codes which treat these elements with varying degrees of complexity, including GALAXEV \citep{bruzual03}, P\'{E}GASE \citet{fioc97, leborgne04}, \citet{maraston05}, \fsps{} \citep{conroy09}, and BPASS \citep{eldridge17}. In some cases, non-stellar effects such as nebular and dust emission can be included separately from the core stellar populations synthesis codes.

The statistical inference framework performs the heavy task of matching complex physical models of galaxy emission to observations. This is usually accomplished through a likelihood function. Many techniques for SED-fitting in the literature thus far have been based on maximum-likelihood optimization, a process sometimes called inversion \citep{walcher11}. The optimization may be through non-linear or, for restricted models, linear methods, or a mix of both \citep{heavens00, cid-fernandes05, ocvirk06a, tojeiro07, koleva08}.
These techniques are popular in part because they are fast and simple. However, many of the likelihood spaces in SED-fitting are both highly non-Gaussian and ill-conditioned such that a small change in the input (e.g., noise in galaxy photometry) can lead to a large change in the output (e.g., star formation histories) \citep{ocvirk06a}. In this case the pure maximum-likelihood estimation and associated inversion techniques cause large noise amplification, resulting in unstable maximum-likelihood solutions and causing severe difficulties in accurately assessing the uncertainties. While regularization can mitigate noise amplification, it complicates interpretation. Furthermore, it can be difficult to incorporate important nonlinear model components such as dust attenuation and spectral calibration in a maximum-likelihood framework.

A solution to these problems can be found in Bayesian forward-modeling techniques. These techniques are able to accurately extract complicated, large, and correlated parameter uncertainties from galaxy observations. Early Bayesian codes such as \citealt{kauffmann03,salim07}, CIGALE \citep{burgarella05}, and MAGPHYS \citep{dacunha08}, were successful in this effort. These codes often worked by comparing pre-computed parameter grids or ``libraries'' of SED models to the observations. While likelihood evaluation over a grid can be computationally very fast, grids grow in size exponentially with the number of parameters. As a result grids only allow a relatively small number of parameters or coarse spacing. Furthermore, grids and libraries are large on disk and slow to regenerate, making it difficult in practice to test the effect of the grid assumptions on the inferred parameters.

The success of these early Bayesian forward-modeling codes motivated an expansion of the technique into gridless `on-the-fly' modeling coupled with Markov chain Monte Carlo (MCMC) algorithms to efficiently explore the resulting parameter space. This approach allows priors and fit parameters to be changed quickly on an object-by-object basis, at the cost of longer per-object computational times. This approach was pioneered by GalMC \citep{acquaviva11}, followed later by codes such as BEAGLE \citep{chevallard16} and BAGPIPES \citep{carnall18}. Here we present \pro{}, a modular galaxy stellar populations inference code based on Bayesian forward modeling and Monte Carlo sampling of the parameter space using \fsps{} stellar populations. \pro{} has been in development since 2014 and, in addition to extensive testing on nearby galaxies \citep{leja17,leja18}, has seen early use in modeling the SEDs of dwarf galaxies \citep{pandya18, greco18}, modeling spatially resolved spectroscopy \citep{patricio18, oyarzun19}, in modeling the host galaxies of supernovae and gravitational wave events \citep{blanchard17a, blanchard17b, nicholl17}, and in modeling gravitationally lensed high-redshift quiescent galaxies \citep{ebeling18}.

The paper is structured as follows.  In \S\ref{sec:ingredients} we describe the ingredients of the \pro{} code including the physical galaxy model, modelling of observational effects, and the statistical framework.  We also discuss some general considerations when building models from these ingredients. In \S\ref{sec:mock} we present several demonstrations of \pro{} fits to mock data, including both photometry and spectroscopy. In \S\ref{sec:real} we give examples of \pro{} operating on real spectroscopic and photometric data. In \S\ref{sec:discussion} we discuss lessons learned during the development of \pro{}, and several avenues for future improvement.  Finally, the appendices provide implementation details.  All fits and figures in this paper were made using Prospector version 1.0.0.

\section{\pro\ Ingredients}
\label{sec:ingredients}

The \pro\ code works by forward modeling the observed spectrum and photometry of complex stellar populations given a combination of parameters describing that population and instrumental parameters, computing a likelihood and posterior probability for the model based on a noise model and prior distributions, and Monte Carlo sampling the posterior probability distribution. In this section we describe, in turn, each of the ingredients necessary for this process, ending with some advice for building models from these ingredients.  Readers interested in demonstrations of the code applied to mock and real data can skip to the next sections.

\subsection{Stellar Population Sources}
\label{sec:sps}

In \pro{} the generation of a basic spectrum and photometry is handled by stellar population \emph{sources}.  These source objects take a set of parameters and compute a mass-normalized restframe spectrum.

In this paper we will model both simple and composite stellar populations, describing star clusters and galaxies respectively.  The spectra for these sources in \pro{} are derived from the Flexible Stellar Population Synthesis \citep[\fsps,][]{conroy09, conroy10a} code.  The \fsps{} code is accessed through the \pyfs{}\footnote{https://github.com/dfm/python-fsps} \citep{pyfsps} bindings.  Given a set of stellar population parameters \pyfs{} is used to calculate an spectrum integrated over stellar age and metallicity.  If necessary, new simple stellar populations (SSPs) may be generated by \fsps{} during this computation; otherwise the SSPs are cached by \pyfs{} and only the calculations required to produce the spectrum of the composite stellar population (as well as nebular emission, dust, and other effects) are made, significantly speeding up the calculation.  This allows for the spectrum of a stellar population to be constructed efficiently on the fly; no parameter grids are necessary, except those used internally by \fsps{}.

The \fsps{} code includes detailed stellar populations with a variety of choices for stellar isochrones and stellar spectral libraries, self-consistent absorption and emission by dust, self-consistent nebular emission, and IGM attenuation.  More details about the ingredients of \fsps{}, especially those that have been added or modified since \citet{conroy10a}, can be found in Appendix \ref{apx:fsps}.  Every parameter available in the \fsps{} code is a potential parameter of a \pro{} model, and can either be free or fixed.  This includes parameters that affect the computation of the SSPs like the slope of the IMF, shifts in the temperatures or luminosities of AGB isochrones, and changing blue-straggler fraction, as well as more traditional composite stellar population parameters like stellar metallicity and dust attenuation (including the slope of the attenuation curve and the age dependence of the optical depth.)  Nebular emission parameters may also be varied.  Varying parameters that affect the SSPs will trigger their regeneration, which is computationally expensive.  As will be discussed in \S\ref{sec:ingredients:zred}, for several parameters available in \fsps{} -- including the redshift and spectral smoothing -- the computations are handled by \pro{} externally to \fsps{}.

\pro{} has access to all of the parametric star formation histories (SFHs) built into \fsps{}.  These forms include exponentially declining SFRs, exponentially declining SFRs with a linear rise (delayed-exponential), constant, single age bursts, and a flexible 4-parameter delayed-exponential model which transitions to a linear ramp at late times described in \citet{simha14}.  In addition, \pro{} includes the ability to model linear combinations of these SFHs.  For example, it is possible to model $SFR(t)$ as the sum of an exponentially declining SFH and a constant SFH, each with separate dust obscuration, to represent the stellar population of the bulge and disk respectively.  It is advised that any user-defined SFH should be tested by generating and fitting mock data to ensure that the information content of the data is well-matched to the complexity of the model (\S\ref{sec:model_building}).

Through the use of tabular SFHs in \fsps{} the \pro{} code can also be used to model and fit more flexible or complicated SFHs.  As currently implemented in \pro{} these are so-called ``non-parametric'' SFHs, high dimensional piece-wise constant (or step-function) star formation histories, the fundamental parameters of which are stellar mass formed in each of $N$ bins of lookback time of user defined width. Additional forms (e.g. piece-wise linear, with parameters giving the SFR at particular times) may be included in the future. Non-parametric SFHs are highly flexible and capable of modeling star formation histories with sharp features or unusual shapes which are forbidden by the usual parametric forms.  It is difficult to constrain flexible star formation histories with typical galaxy observations, so the chosen prior is very important \citep[\S\ref{sec:priors}, \S\ref{sec:illustris}][]{leja19a}.  Through careful parameter transformation a number of different priors on the amplitudes are possible; \pro{} is packaged with four families of such priors, but the user may adjust these or explore more complicated priors.

Finally, the modular nature of \pro{} allows for sources other than the ones available through \fsps{} to be used, in principle.  All that is required is a Python object that, given a parameter dictionary and a set of wavelengths and filters, can compute and return the mass normalized rest-frame spectrum.

\subsection{Redshift, Filters, LSFs}
\label{sec:ingredients:zred}

When modeling data, the models need to be ``projected'' into the data space in various ways.  These effects are handled by code that operates on the spectrum returned by any of the \emph{sources} described above. They include redshifting, normalization by the total formed stellar mass and distance dimming, and filter projections.  If required, filter projections are handled by the pure Python code \sedpy\footnote{https://github.com/bd-j/sedpy} \citep{sedpy} where the addition of user-defined filters is straightforward.

To model observed spectra the normalized, redshifted spectrum must be broadened to match the observed line-spread function (LSF) including physical and instrumental broadening. \pro{} includes code for fast spectral broadening in velocity or wavelength space using fast Fourier transforms (FFTs).  The broadening kernels are currently limited to Gaussians, but more complex kernels may be added in the future.  Broadening with FFTs uses the analytic Fourier transforms of the broadening kernels \citep[e.g.][]{cappellari17}. A detailed discussion of the \pro{} treatment of broadening due to the line-of-sight velocity distribution, the instrumental resolution, and considering the library resolution is given in Appendix \ref{apx:smoothing}.
At the time of writing the highest resolution models available in \fsps{} and hence in \pro{} are the empirical MILES library, with FWHM $\approx 2.5$\AA{} from 3525-7400\AA{}.  Modeling even moderate resolution spectroscopic observations of rest-frame wavelengths outside of this region, or high resolution rest-frame optical data, will require new stellar libraries.  Alternatively, the data may be smoothed to a lower resolution, but this induces correlated noise that would need to be accounted for in the likelihood (see Appendix \ref{apx:noisemodel}).
Finally, because the kinematics of the gas and stars may be different, the nebular lines may need a different broadening and redshift than the stellar continuum. \pro{} has this functionality: see Appendix \ref{apx:nebular} for details.

Adjustments to the wavelength calibration can be modeled using polynomials, following \citet{koleva09}.  Additional modeling of sky emission and transmission is under development.

\subsection{Spectrophotometric Calibration}
\label{sec:ingredients:cal}

The treatment of spectro-photometric calibration and the uncertainty thereon is of critical importance when modeling spectra, and especially when combining spectroscopic and photometric information (\S\ref{sec:likelihood}, \S\ref{sec:mock_par_spec}, Appendix \ref{apx:noisemodel}.) It is common when modeling spectroscopy to remove the sensitivity of the inferred physical parameters to the continuum shape, due to the difficulty of accurate spectro-photometry \citep[e.g.]{koleva09, cappellari17}. Another way to think of this is that the information in the continuum shape of the spectroscopic data should not flow to the physical parameters of interest but rather to nuisance parameters describing the uncertain spectral calibration. Here we describe approaches to the modeling of spectrophotometric calibration that are available in \pro{}.  We assume throughout that the spectra and the photometry probe the same stellar populations, i.e. that there are not substantial color-dependent aperture corrections.  The overall scaling of the spectroscopic calibration vector incorporates any gross aperture corrections, such that normalization sensitive parameters (e.g. stellar mass) correspond to the population probed by the integrated photometry.

The most straightforward approach is to include a parameterized function of wavelength or pixel, typically a polynomial, that multiplies the model spectrum to produce the observed spectrum. One can then fit for the parameters of this calibration function simultaneously with the physical parameters \citep[e.g.][]{lee11, becker15}. The calibration parameters can be marginalized over when considering physical parameter estimates. It is straightforward to include priors on these parameters to incorporate external knowledge about the spectrophotometric calibration accuracy.  However, it can be technically difficult to make this simultaneous fit when the number of calibration function parameters is very large (e.g. for a very high order polynomial).
In \pro{} this difficulty can be avoided by \emph{optimizing} the parameters of the calibration function at each model generation. In this case the computational cost is small, but the resulting posterior distributions do not fully incorporate the uncertainties in the spectrophotometric calibration.
It is possible -- for certain prior distributions and calibration functions -- to analytically marginalize over the parameters at each likelihood call, but this option is not yet available in \pro{}.
A different approach that avoids an explicit calibration function is to model calibration effects as co-variant or correlated noise, which is discussed further in Appendix \ref{apx:noisemodel}.

The parameterized calibration function available by default in \pro{} is a Chebyshev polynomial of the wavelength vector of user defined order $N_p$.  The scale of the calibration features that can be captured with this model is $\sim (\lambda_{\rm max} - \lambda_{\rm min}) / N_p$.  Typically optimization of the polynomial will serve most users' needs.  When no photometry is available the parameters of the calibration vector are unconstrained except by the space of possible continuum shapes allowed by the physical model.  This will lead to unconstrained stellar mass and dust attenuation, as these parameters are typically constrained by the shape and normalization of the continuum.

\subsection{Likelihood, Priors, and Sampling}
\label{sec:likelihood}

\subsubsection{Likelihood}
The basic log-likelihood calculation in \pro{} is effectively a $\chi^2$ calculation for both the spectral and the photometric data.  The two log-likelihoods are then summed.  This gives the following expression for the $\ln$-likelihood:
\begin{eqnarray}\displaystyle
\ln \like(d_s, d_p\given \theta, \phi, \alpha, \beta) &=&
  \ln \like (d_s\given \theta, \phi, \alpha) \\
  && \quad +\, \ln \like (d_p\given\theta, \beta)
  \quad , \nonumber
\end{eqnarray}
where $d_s$ is the spectroscopic data, $d_p$ is the photometric data, the parameters $\theta$ describe the physical model and generally include any of the \pyfs{} parameters, the parameters $\phi$ describe spectroscopic instrumental effects and calibration, the parameters $\alpha$ describe the spectroscopic noise model, and the parameters $\beta$ describe the photometric noise model.

By default we assume known and independent Gaussian flux uncertainties.  Under this assumption the $\ln$-likelihood for the spectroscopic and photometric terms is simple $\chi^2$. More complex noise models are available in \pro{}. These may be necessary to model a variety of instrumental artifacts and are discussed further in Appendix \ref{apx:noisemodel}.

\subsubsection{Priors}
\label{sec:priors}
Before each likelihood call, the prior probability of every free parameter is computed.  A prior probability distribution \emph{must} be specified for each of the parameters that is allowed to vary during a fit.  Several simple distributions are provided including uniform, normal or Gaussian, clipped normal, log-normal, log-uniform, Student's-t, and Beta distributions.  The parameters describing these distributions -- such as the minimum and maximum for the uniform distribution, or the mean and dispersion for the normal distribution -- must be specified as well.  Users may provide their own prior distributions.  It is also possible in principle to specify joint priors on several parameters (e.g. that one parameter be less than another parameter), though in practice this is more easily achieved through parameter transformations (see below).

In the often highly degenerate space of SED modeling, priors can play a strong role in determining the shape of the posterior probability distribution \citep[e.g.][]{carnall19a, leja19a}.  We therefore strongly encourage exploration of the sensitivity of the results from \pro{} to choices of prior, as discussed further in \S\ref{sec:model_building}.

Related to priors, \pro{} offers methods for transforming variables.  This works by labeling some parameters as \emph{dependent on} other parameters via a given transformation function.  This mechanism allows one to sample in transforms of the native parameters (e.g. in $\tau^{-1}$ instead of $\tau$) that may be more convenient, to tie parameters to other, varying, parameters (e.g. to tie the gas phase metallicity to the stellar metallicity), and to avoid joint priors such as limits on population age that are a function of the varying redshift of the model.

\subsubsection{Optimization and Sampling}
\label{sec:sampling}

The posterior probability distribution is sampled using Monte Carlo techniques, and optimization can also be performed (though is not recommended for the reasons described in the Introduction, except as an initial guess for sampling algorithms.)

The primary inference technique used by \pro{} is nested sampling.  Nested sampling \citep{skilling04} is an algorithm based on successive draws from the prior distribution at increasing values of the likelihood.  It uses the change in the effective prior volume of iso-likelihood contours as a function of likelihood to estimate the Bayesian evidence, and the successive draws -- along with their associated \emph{weights} -- can be used to estimate the posterior distribution of the parameters. Because of the way the prior volume is sampled, nested sampling is well-suited to posteriors that are multi-modal (e.g. for photometric redshifts). Furthermore, no optimization is required before sampling begins, and stopping criteria based on estimates of the remaining evidence can be defined. Nested sampling in \pro{} uses the \dynesty{} pure-Python code package \citep{speagle20}, an implementation of algorithms described in \citet{feroz09, higson19}.

Within \pro{} it is also possible to use the ensemble MCMC algorithm of \citet{goodman10}, as implemented in the \emcee{} Python package \citep{foreman-mackey13}.
This algorithm is based on the use of many ``walkers'' in parameter space, the distribution of which is used to construct parameter proposals for the next Monte Carlo step in the chain.

The results of the Monte Carlo sampling are a list of parameter values, which represent (weighted) samples from the joint posterior probability distribution for all the parameters, given the data.  These samples provide a natural and useful tool for quantifying parameter uncertainties (marginalized over other parameters of the model) and the degeneracies between parameters.  Given samples from the posterior probability distribution, it is up to the user to decide what to report.  This may include full set of samples or various projections of the posterior PDF.  It is also possible to report moments or quantiles of the marginalized posterior PDF for each parameter, though care must be taken in the interpretation when the posteriors are very broad and/or non-Gaussian.
For complicated posterior distributions the set of marginalized one dimensional posterior medians may describe a model that is itself very unlikely.  For these reasons it is often best to work with the entire posterior distribution (as approximated by the Monte Carlo samples) for subsequent analyses. For visualizing the model in the space of the data the highest posterior probability sample\footnote{In a gridless search with a finite number of samples, the highest posterior probability \emph{sample} is not the same as the maximum a posteriori solution.} is also useful.

\begin{figure*}
\includegraphics[width=\textwidth]{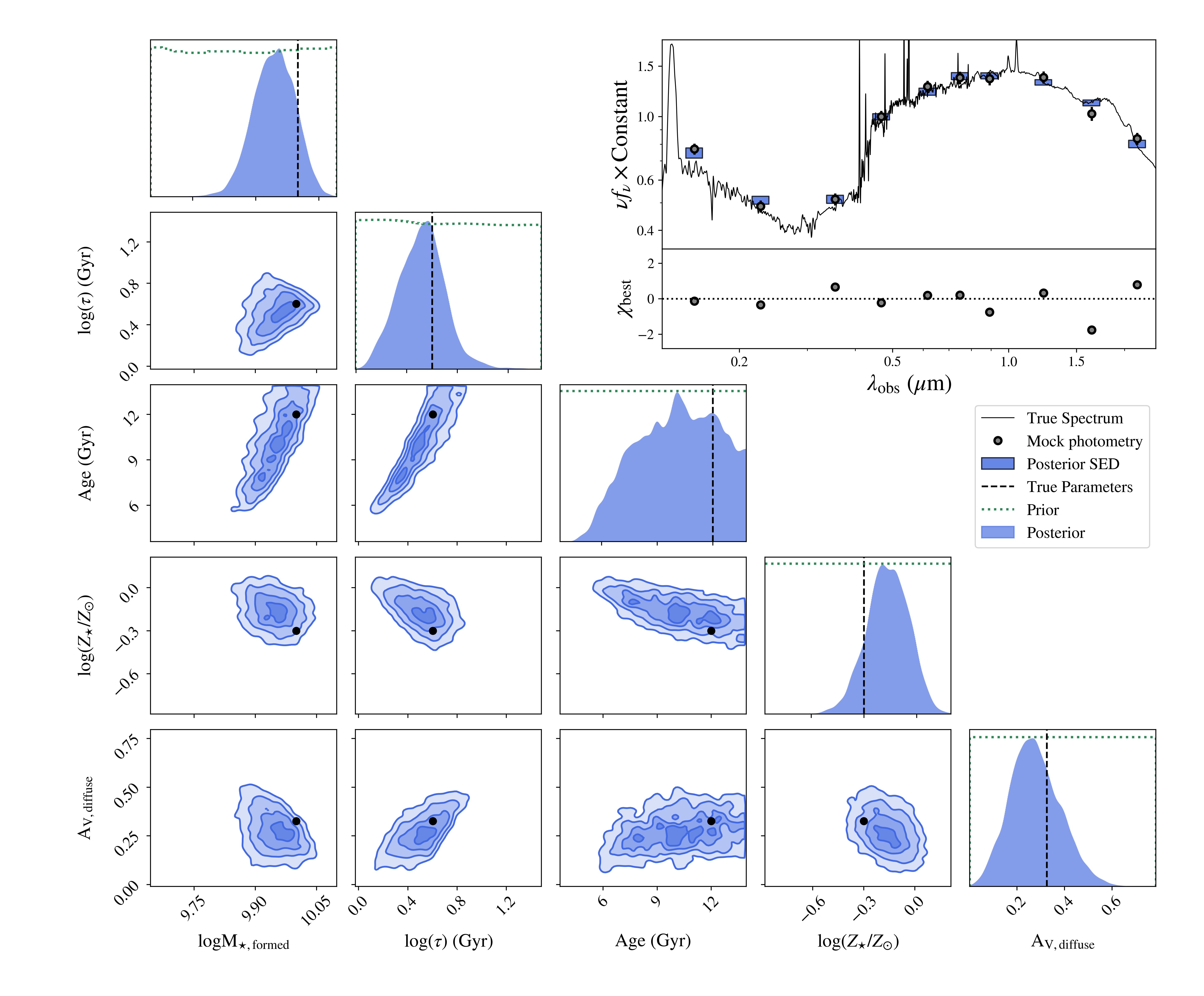}
  \caption{An example of simple parameter inference from a mock UV through Near-IR photometric SED with \pro{}.
  {\it Lower left panels:} Several projections of the posterior PDF for all 5 fitted parameters of a simple delayed exponential SFH model (blue contours and histograms) compared to the true input values (black points, black dashed lines). The one-dimensional projections also show the prior probability distributions (green dotted lines).
  {\it Inset Top:} Mock SED for a delayed exponential SFH.  We show the true input mock spectrum, smoothed to R$\sim 500$ (black line), the mock photometry with S/N$=20$ (orange circles), and the posterior PDFs for the true photometry (blue rectangles).  The height of each rectangle represents the 16th-84th percentile range of the posterior PDF for the true photometry.
  {\it Inset Bottom:} Uncertainty normalized residual ($\chi$) between the mock photometry and the predicted photometry of the most probable Monte Carlo sample (orange).
\label{fig:mock_par_full_posteriors}}
\end{figure*}

\subsection{Model Building}
\label{sec:model_building}

Given all these ingredients in \pro{}, it is the responsibility of the user to define a model that is appropriate for their data and scientific question.
In \pro{} this amounts to choosing a set of parameters, deciding which are free and specifying prior distributions for those, and specifying the values of the remaining fixed parameters.  It may also require defining any parameter transformations to convert from the sampling parameters to the native parameters used to generate spectra and SEDs.  Several examples of this process are presented in \S\ref{sec:mock} and \S\ref{sec:real}.  A significant advantage of on-the-fly model generation is that changing the model definition -- the free parameters, the priors, and the parameterization -- does not require rebuilding large grids of models.

When constructing a model for inference in \pro{}, it is often useful to start with a simple model, and to add complexity as required by the data and scientific question. A first step is to make predictions with a simple model and some fiducial or very approximate parameter values, and to compare this to some representative data.  Once the model seems like it has reasonable components, the next step is to perform a fit and then, crucially, to examine residuals.  This can reveal missing components in the model that were not immediately apparent, such as un-modeled weak nebular emission lines, resolution mismatch, or the presence of hot dust emission. Examining the posterior PDFs can identify priors that are too restrictive or mis-specified.  Finally, with a basic model in hand, one can explore letting parameters that were previously fixed be free, though with informative priors.

There is a natural desire to avoid complexity in the model by limiting the number of free parameters. However, one of the principal advantages of Bayesian inference and lack of grid is the ability to add extra parameters that may affect the SED and \emph{marginalize} over them if they are not scientifically interesting.  Adding such parameters -- the values of which are not extremely well known {\it a priori} but that might affect the SED -- will cause other parameter estimates to become less certain, and thus guard against over-interpreting the data.  If the parameter is not well constrained by the data -- that is, if it does not affect the data in a unique and significant way -- then this will be reflected in the posterior PDFs, which will take the form of the prior for that parameter.  Furthermore, the constraints on other parameters will only be affected if they are somehow degenerate with the new parameter; in this case, \emph{not} including the additional parameter would bias the uncertainties low on these other parameters.  Thus one should not shy away from including poorly constrained parameters that are known to be physically important or to affect the SED, though in such cases it is important to put thought into the adopted prior distribution.  The only drawback to adding parameters is that MCMC sampling can become inefficient as the number of parameters increases.

When the data are only weakly informative about the parameters, as can often occur in SED fitting, the form of the priors can play a strong role in the posterior estimation.  It is often useful to place \emph{informative} priors on some parameters, by which we mean here priors that are not uniform.  If one has external information about the plausible values that a parameter might take -- for example that the shape of the attenuation curve is most likely to be a certain value but that there is scatter -- then it is best to use that in the form of a prior.  This is preferred to the typical approach of fixing the parameter to the single most-likely value: that effectively constitutes an extremely strong prior and will often result in overly-confident posteriors.

However, the importance of priors can also raise concerns about the interpretation of the results -- did you actually learn anything from fitting your data, or are the results dominated by the prior assumptions?  When this is a concern, we recommend testing the sensitivity of the results to the form of the prior.  This can be done most directly by running fits with a different but still plausible prior, and seeing if and how the results change.  It can also be useful to generate SEDs or other derived quantities by directly sampling the chosen prior distribution and inspecting the distribution of those quantities to see what values would be preferred or allowed by the model.

\section{Demonstrations with Mock Data}
\label{sec:mock}

In this section we provide demonstrations of some of the capabilities of the \pro{} code for fitting galaxy spectra and/or photometry using mock data. These demonstrations include the combination of photometric and spectroscopic information in a single fit, the inference of flexible, many-parameter SFHs, and the constraints provided on a complex model with different combinations of photometric data, including a single optical photometric point.

Throughout these demonstrations with mock data we use the MIST isochrones \citep{choi16} and the MILES spectral library \citep{sanchez-blazquez06} within FSPS to construct both the mocks and the models with which they are fitted. Because we are fitting mock data with the same stellar population models used to generate that data, the results of this section allow us to demonstrate and test the \pro{} methodology and examine the constraints that can placed in the limit of perfect stellar population models.

The demonstrations in this section are not intended as exhaustive explorations of the information content of different kinds of data and the suitability of different kinds of models, but rather provide examples of the capabilities of the \pro{} code for fitting different kinds or combinations of data with a variety of galaxy SED models.

\subsection{A Simple Parametric SFH Model}
\label{sec:mock_parametric}

Since the pioneering work of \citet{tinsley80} it has been common to fit observed SEDs to galaxy models where the SFH is represented by a relatively simple functional form characterized by a few parameters, $SFR(t) =f(t,\theta)$.  The commonly-used exponentially declining star formation history arises in closed-box evolutionary models with a constant star formation efficiency.  In addition to the exponentially declining and the related delayed-exponential SFHs, recent studies have proposed a variety of functional forms thought to be more or less well matched to the SFHs of real galaxies \citep{gladders13, simha14, diemer17, carnall18}. Parametric SFH models attempt to describe the potentially complex evolution of galaxies with only a few numbers.  This is very desirable for limiting the size of SED grids and making any Bayesian inference with MCMC more tractable.

Here we present a fit to mock data with a parametric SFH model.  This provides a comparison point for the extensive literature using parametric SED models, and the simplicity of the model allows us to focus on different aspects of the \pro{} fitting.  For all of the fits presented in this section we use a delayed-exponential SFH where the SFR $\psi$ as a function of lookback time $t_\ell$ is given by
\begin{eqnarray}
 \psi(t_\ell) & \propto & \frac{\tage-t_\ell}{\tau} e^{-(t_{age}-t_\ell) / \tau} \quad,
               \quad 0 < t_\ell < t_{age}
\end{eqnarray}
with parameters $\tage$ and $\tau$. While a number of treatments of dust attenuation are available through \fsps{} (see Appendix \ref{apx:fsps}), we model the dust attenuation with a simple dust screen affecting stars of all ages equally, and specify a fixed wavelength dependence of the optical depth based on \citet{kriek13}.  The normalization of this attenuation curve -- the optical depth at 5500\AA{} -- is left as a free parameter. We also include as a free parameter a single metallicity for all stars.  For simplicity we tie the metallicity of the nebular emission model (Appendix \ref{apx:fsps}) to this stellar metallicity, but this is not required in general. The ionization parameter $U$ of the nebular emission is left as a free parameter.  Finally, the overall normalization of the SFH is left as a free parameter, corresponding to the total mass of stars \emph{formed} or the integral of the SFH.  We include dust emission in the model, based on the 3-parameter models of \citet{draine07} (see Appendix \ref{apx:fsps} for details), but we do not fit for these parameters, as we do not include sufficiently red data to be sensitive to these parameters.  Unless otherwise stated, values of the fixed parameters used to generate the mock data are the same as the parameters used to fit the mock data.

The modeling of redshift is described and treated separately in the fits below.  However, the redshift of the mock is set to 0.1, and the prior distribution on $\tage$ does not allow solutions older than the age of the universe at this redshift. The values of the other mock parameters (and the priors used when modeling the SED) are given in Table \ref{tbl:mock_parameters}.

\begin{deluxetable}{lll}[t!]
\tablewidth{0.43\textwidth}
\tablecaption{Summary of parameters and priors used in mock demonstrations\label{tbl:mock_parameters}}
\tablehead{
\colhead{Parameter} &\colhead{Mock Value} & \colhead{Prior Functions}
}
\startdata
Stellar Mass\tablenotemark{a}  [M$_{\odot}$]  & $10^{10}$ & LogUniform($10^7$, $10^{13}$)   \\
Age $t_{\rm age}$ [Gyr]                       & 12        & Uniform($0.1$,$t_{\rm univ}$\tablenotemark{b})\\
$\tau_{\rm SF}$ [Gyr]                         &  4        & LogUniform($0.01$,$100$) \\
$z$                                           & 0.1       & $\norm(0.1$,$0.001)$ \\
$\log(Z_\star/Z_\odot)$                       & $-0.3$      & Uniform($-2.0$,$0.2$) \\
$\tau_{5500}$                                 & 0.3       & Uniform($0$,$4$) \\
$\log U$                                      & $-2$        & Uniform($-4$,$-1$) \\
$\sigma_v$\tablenotemark{b} [$\kms$]          & 200       & Uniform($30$,$400$) \\
\enddata
\tablecomments{Uniform($x$,$y$) indicates a uniform prior from $x$ to $y$, while $\norm (\mu,\sigma)$ indicates a normal distribution for the prior and LogUniform($x$, $y$) indicates distribution that is uniform in the log of the parameter in the parameter range ($x$, $y$)}
\tablenotetext{a}{This is the integral of the SFH.}
\tablenotetext{b}{The maximum age is set by the age of the universe at the model redshift in a WMAP9 cosmology.}
\tablenotetext{c}{Velocity dispersion is only included as a model parameter when fitting spectra}
\end{deluxetable}

\begin{figure*}
\includegraphics[width=\textwidth]{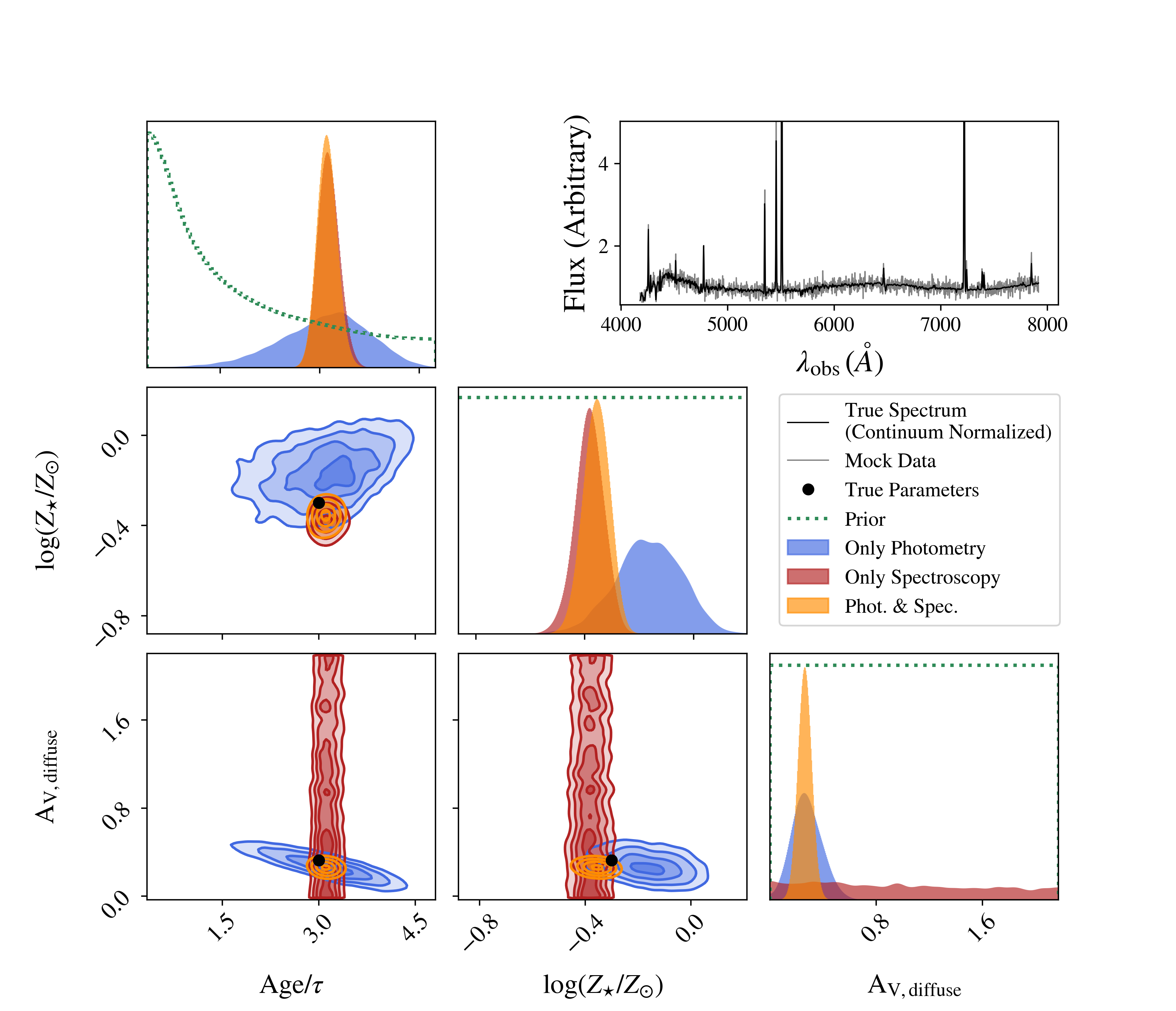}
  \caption{Comparison of parameter inference from photometry, continuum normalized spectroscopy, and the combination.
  {\it Left panels:} Several projections of the posterior PDF for selected parameters of a parametric SFH model. The parameters inferred from photometry only (blue shaded regions), continuum normalized spectroscopy only (red shaded regions), and photometry plus continuum normalized spectroscopy (orange shaded regions) are compared to the true input parameters (black points, black dashed lines). The prior probability distributions, scaled by an arbitrary factor, are shown in the one-dimensional projections (green dotted lines).
  {\it Top right panel:} The true spectrum, roughly continuum normalized as described in the text, is shown (black) along with the noisy mock spectrum (gray)
\label{fig:mock_par_posteriors}}
\end{figure*}

\subsubsection{Inference from Photometry}
\label{sec:mock_par_phot}
In this section we demonstrate inference of the parameters of the simple model described above from only the mock photometric data. The mock photometry is generated using the same model used for the inference, with parameters given in Table \ref{tbl:mock_parameters}.  We generate photometry in the filters of several large area sky surveys including {\it GALEX} FUV and NUV, SDSS $ugriz$, and the 2MASS $JHK_s$ bands.  We assume uncorrelated uncertainties of 5\% in the photometry in each band, and add a realization of the noise to the mock photometry.

For this demonstration we assume, when fitting the photometry, that the redshift is constrained to the true value to within $0.001$. That is, we include the redshift as free parameter, but with a very narrow prior distribution as might be expected from low S/N grism spectra, yielding a model with 7 free parameters. We then use \pro{} to explore the posterior probability distribution of the free parameters for this model.  For the inference we use nested sampling, with default parameters.
In Figure \ref{fig:mock_par_full_posteriors} we show the mock spectrum and the mock photometry, the latter perturbed by a realization of the noise.  We also show the inferred values for the true photometry as shaded boxes with heights indicating the 16th - 84th percentile range of the posterior probability. Uncertainty-normalized residuals for the highest-probability posterior sample are also shown.  Examining such residuals is a key step in evaluating the performance of any model and identifying any deficiencies that require additional model components or flexibility to adequately describe the data.

In Figure \ref{fig:mock_par_full_posteriors} we also show projections of the posterior probability distribution for several of the free parameters (redshift and the nebular ionization parameter are not constrained by the data, and the posterior distributions are indistinguishable from the prior).  The input mock parameter values are also shown in black, and the adopted prior probabilities are indicated in the one-dimensional projections.  The posterior PDFs for a number of the parameters are quite broad.  Notably there is a strong degeneracy between $t_{\rm age}$ and $\tau_{\rm SF}$ such that while neither is well constrained, their ratio is reasonably well constrained.  The ratio $t_{\rm age}/\tau$ is closely related to the specific star formation rate.
The well-known dust-age-metallicity degeneracy is mapped in complete detail.

It is worth pointing out that the peak of the marginalized posterior probability distributions are not generally centered on the true values.  This is a common and expected occurrence, caused in part by the perturbation of the mock fluxes by a realization of the noise, and in part by the complicated shape of the posterior in multidimensional parameter space combined with the adopted priors. This highlights the potential drawbacks of simply using the maximum a posteriori parameters.

\subsubsection{Incorporating Information from Spectroscopy}
\label{sec:mock_par_spec}
We now turn to the inference of mock parameters from mock spectroscopic data, both alone and in combination with photometric data. We generate a mock spectrum from the model, smooth the mock spectroscopy by an additional velocity dispersion of $\sigma_v=200 \, \kms$ and assume an instrumental dispersion of 2\AA~ per pixel from rest-frame 3800\AA -7100\AA~ (corresponding to the region of the spectrum covered by the high-resolution empirical MILES stellar library). The applied physical velocity smoothing dominates over the intrinsic resolution of the MILES stellar library, but for clarity and simplicity we also assume that the instrument has the same line-spread function as the MILES library redshifted to $z=0.1$ ($\sim 2.75$\AA~ FWHM). See Appendix \ref{apx:smoothing} for approaches to data where the instrumental or library broadening is a significant contributor to the final line-spread function of either the data or model. We assume 10\% uncertainties in the flux of each pixel, and a realization of this noise is added to the data.
To mimic realistic observations where the spectrophotometric calibration may not be known, and to highlight that we are not using information from the continuum shape,  we divide the mock spectrum by a 6th-order polynomial fitted to regions of the spectrum free from strong absorption or emission lines.

To model the continuum-normalized mock spectrum we include all free parameters used in the fit to the photometry (Table \ref{tbl:mock_parameters}), and we also include the galaxy velocity dispersion as a free parameter.  To account for any uncertainty in the spectroscopic calibration, we use the method of multiplying the model spectrum by the maximum-likelihood polynomial that describes the ratio between the spectroscopic data and the model before computing the likelihood of the data.  We use a 12th-order polynomial that is determined at each likelihood call. By including this polynomial multiplication we do not use information from the spectral continuum shape; that information is supplied by the photometry if it also being fit.

In Figure \ref{fig:mock_par_posteriors} we compare the posterior PDFs for several model parameters inferred from the continuum normalized spectroscopy alone (in maroon contours) to those inferred from the photometry alone (blue contours, previous section.)  Given the strong degeneracy between $t_{age}$ and $\tau$ when fitting to photometry alone, we instead show the posterior PDF for the ratio $t_{age} / \tau$.  In this comparison we see that the posterior PDF for the dust attenuation is much broader when inferred from continuum normalized spectroscopy than when inferred from the photometry, and is essentially the same as the prior.  This is expected since all the continuum shape information in the data is absorbed by the multiplicative polynomial.  In other words, the dust attenuation is completely degenerate with the spectrophotometric calibration, which we have assumed to be completely unknown.  The stellar mass is similarly unconstrained. The metallicity is tightly constrained by the spectroscopy; this is not surprising, as in addition to the metallicity information provided by absorption lines in the stellar continuum we have tied the nebular emission model to the stellar metallicity and include the emission lines in the fitted data.

Figure \ref{fig:mock_par_posteriors} shows that broadband UV through NIR photometry and continuum normalized spectroscopy carry different information about about the model parameters.  It is natural then to combine the constraints from each.  We do this by simultaneously modeling and fitting the photometry and spectroscopy using the combined likelihood described in \S\ref{sec:likelihood} \footnote{This could be accomplished by simply multiplying the posterior probabilities from the separate fits to photometry and spectroscopy, but the advantages of joint fitting include more efficient sampling of the posterior probability distribution and a guarantee of sufficient samples in the region of overlap.}. Because we have included a polynomial calibration or normalization vector when modeling the spectra, the information about the continuum shape and any associated parameter is entirely supplied by the photometric data.  If we did not include this uncertainty in the spectroscopic calibration then the much larger number of spectroscopic points with comparable S/N to the photometric data would overwhelm any information in the photometry, unless two models with equally likely spectra somehow had significantly different photometry.  This latter situation would be likely occur only if the SFH and dust model where very flexible, or the spectra were of much lower S/N than the photometry.

In Figure \ref{fig:mock_par_posteriors} we show the posterior probability distribution functions of the selected parameters obtained from a simultaneous fit to the photometry and the spectroscopy (gold contours).  For all parameters these are narrower than the posterior PDFs obtained from either photometry or spectroscopy alone.  Interestingly, while the continuum normalized spectroscopy places no constraint on the dust attenuation by itself, its combination with photometry results in even tighter constraints on the attenuation than from the photometry alone.  This is because the precision on other parameters such as $\feh$, which are partially degenerate with dust attenuation in the photometry alone, is increased by inclusion of the spectroscopic data.

\begin{figure}
\includegraphics[width=0.5\textwidth]{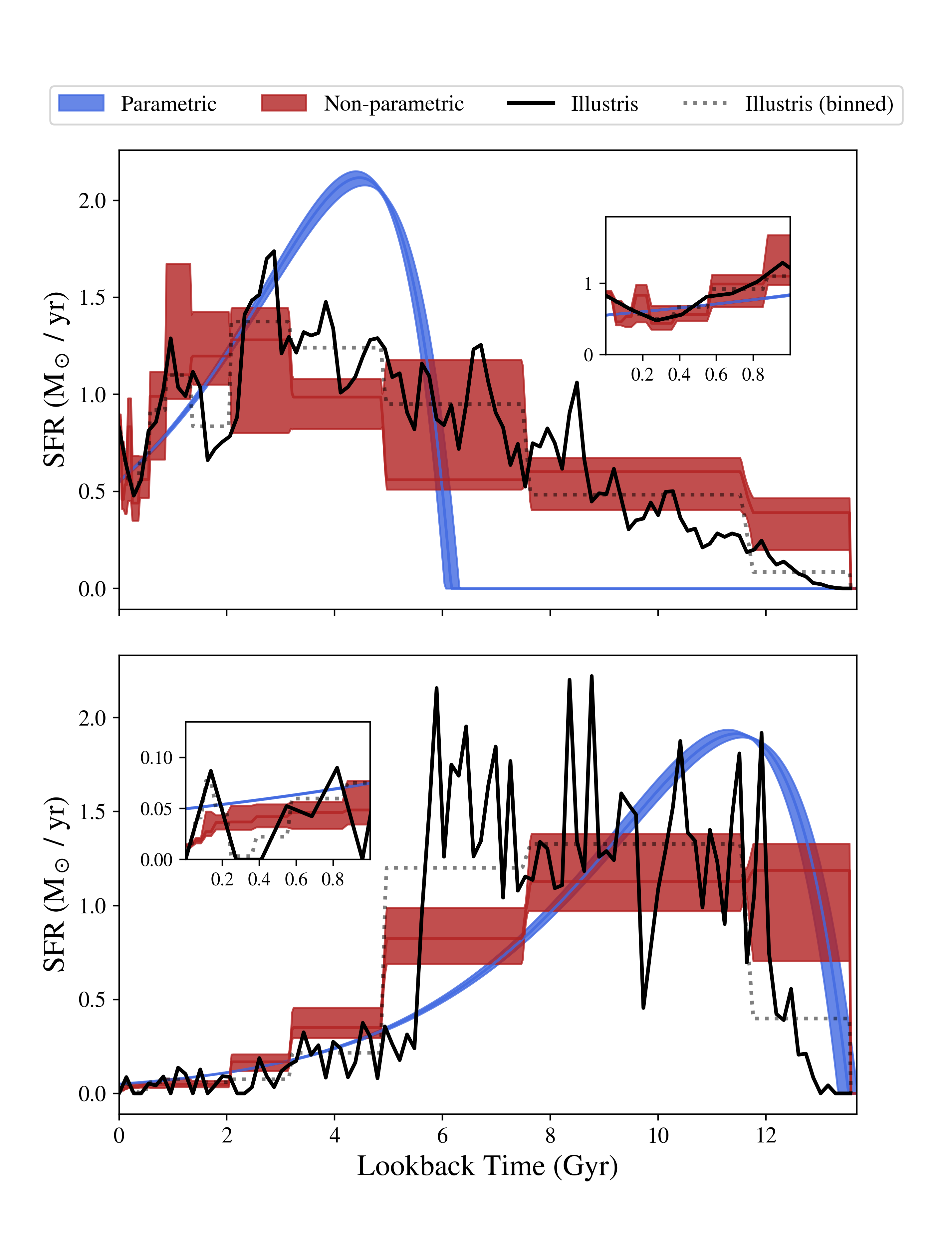}
  \caption{Posterior probabilities for SFHs inferred from S/N$\sim$100 optical spectra with no nebular emission and perfectly known spectrophotometric calibration.  The mock data is generated from an Illustris SFH (black).  The shaded regions show the 16th-84th percentile credible interval for the inferred parametric SFH (blue) and non-parametric SFH (maroon).  The non-parametric SFH uses 14 bins with a continuity prior and the average input SFR in the same bins is shown as dotted gray lines.  The insets in each panel show the recent SFH ($< 1$ Gyr) in more detail.
\label{fig:nonpar_mock_sfh}}
\end{figure}

\subsection{Inferring the SFHs of Illustris galaxies}
\label{sec:illustris}
The assumption of a particular functional form for the SFH of a galaxy amounts to an extremely strong prior on that SFH. However, it is unclear how well real galaxies are represented by these parametric models \citep[e.g.]{gallagher84}.  Deviations from the assumed forms affect the accuracy of derived parameters \citep[e.g.,][]{lee09, pforr12, wilkins12, pacifici15, lower20}, as well as estimates of the uncertainty on the SFH \citep{carnall19a, leja19a}.

One way to address the complex SFHs in real galaxies is to increase the flexibility of the model SFHs.  Flexibility has been introduced previously by adding additional components (i.e. recent bursts), by choosing more flexible forms for the parametric SFH \citep[e.g.,][]{salim07, carnall18}, or using a library of SFHs based on semi-analytic or hydrodynamic simulations \citep[e.g.,][]{dacunha08, pacifici13, iyer19}.  These approaches can require large library sizes to densely sample the prior SFH parameter space, and also incorporate the simulation as an SFH prior.

Another possibility is to use piece-wise constant or $\delta$-function SFHs where the mass in each temporal bin becomes a parameter of the model. These so-called \emph{non-parametric} -- actually many-parameter -- models have become the standard in SFH inference from (semi-) resolved stellar color-magnitude diagrams \citep[e.g.,][]{dolphin97, harris01, weisz11, cook19} and have been used in maximum-likelihood ``inversion'' codes for unresolved spectra \citep[e.g.,][]{cappellari04, cid-fernandes05, ocvirk06a, tojeiro07, koleva09}.  In principle any basis function can be used \citep[e.g.,][]{iyer17} though the piece-wise constant basis has the advantage of being readily interpreted.

\pro{} supports non-parametric models using a piece-wise constant basis (\S\ref{sec:sps}).  In contrast to maximum-likelihood techniques, \pro{} allows for the exploration of the full posterior PDF for the component amplitudes. Inference with this kind of non-parametric SFH was explored in detail by \citet{leja19a}, where a number of different priors on the amplitudes of each bin were considered.  A prior based on the ratio of the amplitudes in adjacent bins has some desirable qualities, and we adopt such a ``continuity'' prior in this section.
 In this model the sampled parameters are the log of the ratio of the SFRs in adjacent time bins and the log of the total stellar mass.  The mass $M_i$ formed in each time bin -- the native parameters used to generate the spectrum -- depend on these sampled parameters through a transform:
\begin{eqnarray}
 \log r_j & = & \log (s_j / s_{j+1}) \\ \nonumber
  s_0 & = & 1 \\ \nonumber
  s_i & = & \frac{s_0} {\prod_{j=0}^{j < i} r^{-1}_j}\\ \nonumber
 m_i & = & s_i \, \Delta t_i \\ \nonumber
  M_i & = & \, m_i \, \frac{10^{logmass}} {\sum_i m_i} \quad ,\nonumber
 \end{eqnarray}
where $\Delta t_i$ is the width of of the $i$th bin of lookback time, in years. We place a Student's $t$-distribution prior on each of the $\log r_j$ parameters, centered on 0 (i.e. constant SFR); the Student $t$ distribution is similar to a Gaussian prior but with more probability in the tails.  The effect of this prior is that in the absence of a constraint from the data the bins default to a constant SFR.

To demonstrate the capabilities of \pro{} to infer both parametric and non-parametric SFHs we construct mock data based on several galaxies from the Illustris cosmological hydrodynamic simulation of galaxy formation \citep{illustris,diemer17}. The galaxies were selected to have sharp variations in both the recent and the extended star formation history. This selection provides the greatest challenge for the inference machinery, and are also weighted against by the chosen prior, ensuring that any agreement is truly driven by the data.
These SFHs have a temporal resolution of 0.136 Gyr (i.e. $t_{univ} / 100$), and we linearly interpolate between the provided SFRs to build the mock spectrum. We assume a single stellar metallicity and a simple foreground dust screen.  The mock spectra are generated for a redshift $z=0.1$ and smoothed by a velocity dispersion of $\sigma_v = 100\kms$. As for the mocks in \S\ref{sec:mock_parametric} we assume an observed frame instrumental line-spread function equivalent to the MILES spectral library at the assumed redshift.  We assume a S/N of 100 for each pixel.  For simplicity we do not include nebular emission.
We infer the SFH from the mock spectra with both parametric and non-parametric SFH models.  For both models we simultaneously infer the SFH, the stellar metallicity, the dust attenuation optical depth for a simple foreground screen, the stellar velocity dispersion, and the redshift.  To focus the analysis on the physical information provided by ideal data we assume that the spectrophotometric calibration is perfect.

For the parametric SFH we use the same delayed exponential model as in \S\ref{sec:mock_parametric}.  For the non-parametric SFH model we adopt 14 bins approximately evenly spaced in $\log t_\ell$.  This choice of the number of bins and their exact widths is driven by experimentation, the difficulty in sampling a very high-dimensional parameter space when using an extremely large number of bins, and considering the degree of difference between the spectra in adjacent bins.

In Figure \ref{fig:nonpar_mock_sfh} we show the input Illustris SFHs and the posterior PDFs for the SFHs inferred with both a parametric and non-parametric model.  In these examples we find that the parametric model will tend to reproduce the shape of the very recent SFH -- which often accounts for the stars that dominate the optical light -- even at the expense of reproducing the shape of the SFH at earlier epochs if that is not allowed by prior.  In contrast the flexibility of the non-parametric SFH model is better able to recover the shape of the recent and older SFH.  However, the influence of the adopted prior on SFR ratios is remains evident, though at a lower level - the recovered non-parametric SFH will tend towards approximately constant SFR when the data are not constraining, and obviously changes in the SFR within a given temporal bin are unrecoverable.  We also find that the constraints on the SFH when using a parametric model are too strong when compared to the non-parametric SFH, and less able to encompass the true range of SFHs consistent with the data (including the true input mock SFH.)  Finally, these differences in the inferred SFHs also affect inferred stellar masses, mass weighted stellar ages, and star formation rates.  For more discussion on these points we refer the reader to \citet{carnall19a}, \citet{leja19a}, and \citet{lower20}.

\begin{figure*}
    \animategraphics[width=\textwidth,controls=on,type=png,palindrome]{1}{nband}{1}{7}
    \caption{
    Change in posterior PDF for the SED and parameters of a complex model as the number of photometric bands increases. The model includes a continuity SFH with 6 temporal bins, a complex dust attenuation model, dust emission parameters, and AGN contribution to the MIR.  There are 16 parameters total.  Nebular emission is also included. The model is fit to a varying number of mock photometric data points, indicating how the model becomes more tightly constrained with the addition of multi-wavelength data (animation).
    {\it Left panel:} The posterior prediction for the SED (blue shaded region, indicating the 16th-84th percentile range for the flux at every wavelength), resulting from a fit to photometry in the indicated bands (orange circles) and compared to the true input SED (black line).
    {\it Right panels:} The marginalized posterior (blue shaded regions) and prior (dotted green lines) probability distributions for selected parameters of the model, compared to the true input parameters (dashed black lines).  Several of the parameters shown are derived from the fundamental fitted model parameters, see text for details.  We show parameters of the stellar population (top row), the dust attenuation model (second row), the dust emission model (3rd row) and the AGN parameters (bottom row).
    \label{fig:nband}
    }
\end{figure*}

\begin{deluxetable}{lll}[ht]
\tablewidth{0.43\textwidth}
\tablecaption{Summary of parameters and priors used in the complex model \label{tbl:nbands_parameters}}
\tablehead{
\colhead{Parameter} &\colhead{Mock Value} & \colhead{Prior Functions}
}
\startdata
\cutinhead{SFH}
$\log M_\star$\tablenotemark{a}  [M$_{\odot}$]  & 10        & Uniform($7$, $13$)   \\
$\log r_i$\tablenotemark{b}                     & 0         & StudentT($0, 0.3, 2$) \\
$\log(Z_\star/Z_\odot)$                         & -0.3      & Uniform($-2.0, 0.2$) \\
\cutinhead{Dust Attenuation}
$\tau_{\rm 5500,diffuse}$                       & 0.5       & $\norm_c(0.3, 1.0, 0, 4)$ \\
$\tau_{\rm young}/\tau_{\rm diffuse}$           & $1$       & $\norm_c(1, 0.3, 0, 1.5)$ \\
$\Gamma_{\rm dust}$                             & 0         & $\norm_c(0, 0.5, -1, 0.4)$ \\
\cutinhead{Nebular Emission}
$\log U_{\rm neb}$                              & $-2$      & Uniform($-4, -1$) \\
\cutinhead{Dust Emission\tablenotemark{c}}
$U_{\rm min, dust}$                             & $2$       & Uniform($0.1, 25$) \\
$Q_{\rm PAH}$                                   & 1         & Uniform($0.5, 7$) \\
$\gamma_{\rm dust}$                             & $10^{-3}$ & LogUniform($10^{-3}, 0.5$) \\
\cutinhead{AGN\tablenotemark{d}}
${\rm f}_{\rm AGN}$                             & $0.05$    & LogUniform($10^{-5}, 3$) \\
$\tau_{\rm AGN}$                                & 20        & LogUniform($5, 150$) \\
\enddata
\tablecomments{Uniform($x$,$y$) indicates a uniform distribution from $x$ to $y$, while $\norm_c (\mu,\sigma, x, y)$ indicates a normal distribution, clipped to the range $(x, y)$ and LogUniform($x$, $y$) indicates distribution that is uniform in the log of the parameter in the parameter range ($x$, $y$).}
\tablenotetext{a}{This is the integral of the SFH.}
\tablenotetext{b}{$r_i$ is the ratio of the SFR in temporal bin $i$ to that in the adjacent bin. There are 5 such parameters that describe the SFH.}
\tablenotetext{c}{Parameters for the \citet{draine07} dust emission model; see Appendix \ref{apx:fsps}.}
\tablenotetext{d}{Parameters for the \citet{nenkova08a} AGN torus emission model; see Appendix \ref{apx:fsps}.}
\end{deluxetable}

\subsection{A Complex Model: The Effect of More Data}
In this section we explore how the inference of a complex SED model is affected by increasing the number of photometric bands used to constrain it. The inferences we have shown to this point have been of relatively simple models, with few parameters. However, more complex models may be required to adequately describe the increasing quality of observations.  Furthermore, making robust inferences from even a small number of low quality data -- and accurately representing the uncertainties on the underlying parameters -- requires modeling all the physical properties that might affect that data.  For example, there is evidence that the effective attenuation curve in galaxies is not fixed, so any quantities inferred assuming a fixed attenuation curve will tend to have underestimated uncertainties.

One reason to make a number of simplifying assumptions about the relevant physical ingredients is that generating and storing grids for high dimensional parameter spaces is computationally challenging. One of the most important features of on-the-fly model generation with Monte Carlo sampling of the posterior is the ability to infer a larger number of physical parameters than can be considered in grid based approaches, since models are only generated in regions of high probability as needed, instead of over the whole parameter space.

The model we explore is described by a non-parametric SFH with 6 temporal bins, a single stellar metallicity for all stars, a dust screen attenuation curve affecting all stars and described by a free power-law index and a V-band optical depth, an extra attenuation towards young stars described by a power law attenuation curve with fixed power-law index, nebular emission described by the gas-phase metallicity and ionization parameter, AGN dust torus emission described by 2 parameters \citep[\ref{apx:fsps}]{nenkova08a, leja18} and dust emission described by the 3 parameters of the \citet{draine07} model.  In total this model has 16 free parameters, listed in Table \ref{tbl:nbands_parameters}.  We assume a known redshift, $z=0.1$.  We adopt informative priors on many of these parameters, particularly the dust attenuation parameters and the AGN parameters.  These prior distributions favor typical dust attenuation curves and disfavor large AGN contributions to the mid-IR.

We generate broadband photometry in standard UV-FIR filters for a mock SED generated for a particular value of the model parameters.  We then infer the model parameters using increasingly large subsets of the photometry, beginning with a single SDSS $r$ band data point.  The other filter sets add SDSS optical bands, then 2MASS near-IR bands, then GALEX UV bands, then WISE mid-IR bands, and finally Herschel far-IR bands.

In Figure \ref{fig:nband} we show the resulting posterior distributions for parameters of the model and for the SED. To summarize the inferred SFHs we derive and present the recent SFR and the mass-weighted stellar age of the population. We see that for a small number of photometric bands the posteriors of nearly every parameter are determined by the prior distribution.  Nevertheless, even with a single photometric point we are able to obtain a (poor) constraint on the total stellar mass.  For this particular mock SED the posterior PDF is shifted to slightly larger stellar mass than the true value. This is largely because the true dust attenuation -- which is entirely unconstrained except by the prior -- is at the lower end of the prior distribution.  Therefore, the prior (and posterior) distribution for the SED is redder and has larger mass to light ratio than the true SED.  This exercise demonstrates how through the use of informative priors and Bayesian reasoning even a single data point can be informative in the context of very high dimensional model.

As the number of photometric bands increases the posterior constraints on many parameters become stronger and, more importantly, distinct from the prior distribution. Adding even a single additional band leads to a much stronger constraint on the dust attenuation and hence mass-to-light ratio than a single band \citep[e.g.,][]{bell01}. With the full SED nearly every parameter has a posterior PDF different from the prior.

\begin{figure*}
\includegraphics[width=\textwidth]{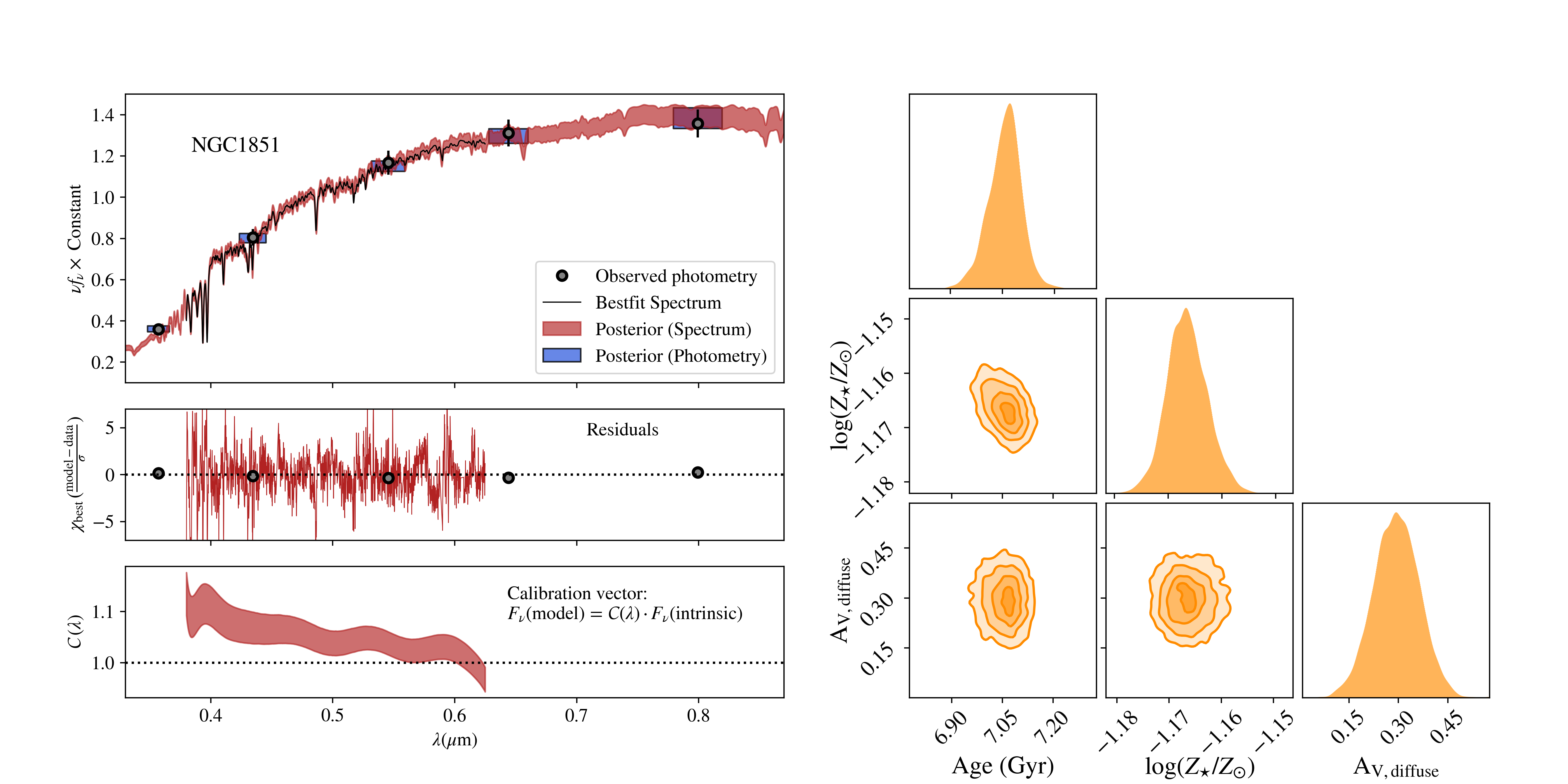}
  \caption{Example of a fit to globular cluster data, showing results for NGC1851.
  {\it Top left panel:} Comparison of the observed photometry (white circles and error bars) to the posterior probability distributions for the true photometry (blue rectangles; height gives the 16th-84th percentile range) and the true intrinsic spectrum (shaded red, showing 16th-84th percentile range). We also show the intrinsic spectrum for the most probable posterior sample (thin black line).  All spectra are smoothed to $R\sim 500$ for display purposes.
  {\it Middle left panel:} Uncertainty normalized residual between the observed spectrum and the model spectrum of the most probable posterior sample (thin red line).  Note that these are normalized by the nominal uncertainties (median S/N$=320$), and do not include the fitted noise inflation term. The uncertainty normalized residuals for the photometry are also shown (white circles).
  {\it Bottom left panel:} The inferred spectrophotometric calibration vector (red shaded region, 16th-84th percentile), defined as the ratio between the observed spectrum and the true intrinsic spectrum.
  {\it Right panels:} Projection of the posterior PDF for selected parameters of the model, inferred from the combination of photometry and spectroscopy. The priors for the parameters shown are uniform.
\label{fig:gc_demo}}
\end{figure*}

\begin{figure}
\includegraphics[width=0.5\textwidth]{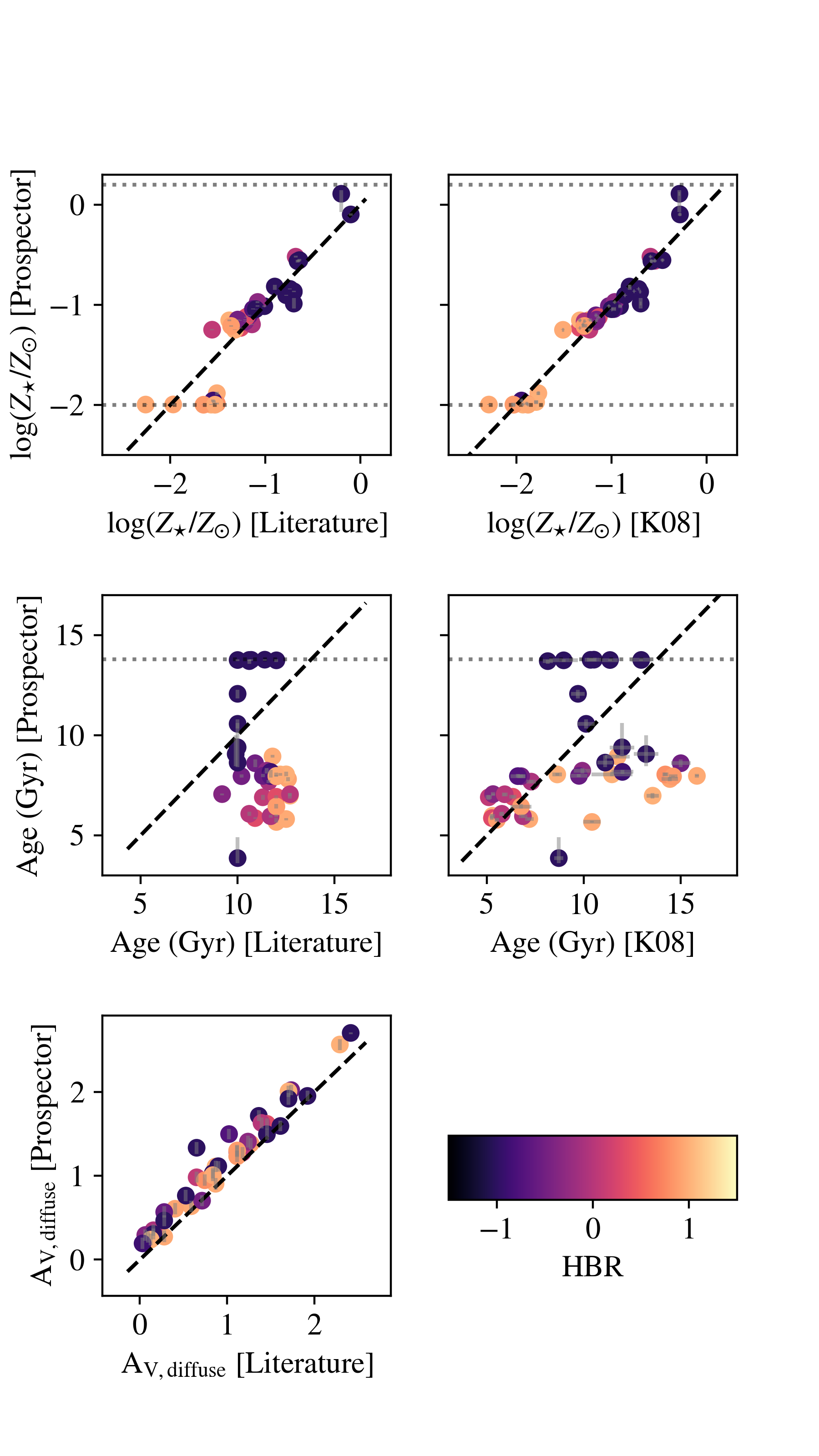}
  \caption{Comparison of globular cluster parameters inferred with \pro{} to literature values and to  values inferred with ULySS by \citet{koleva08}.
  Literature ages are largely from CMD fits, as tabulated in \citet{koleva08}. Clusters without reliable age determinations were assigned an age of 10 Gyr.  Literature metallicities are from high resolution studies of individual stars, again as tabulated in \citet{koleva08}.  Literature extinction is computed from the E(B-V) values in \citet{harris96}. The color coding is by a measure of horizontal branch color (HBR) in each cluster. The bluest or hottest horizontal branches have HBR$\sim +1$.  Prior limits on metallicity and age are indicated as gray dotted lines.
  {\it Top Row:} Stellar metallicity.
  {\it Middle Row:} Globular cluster age.
  {\it Bottom panel:} Dust extinction.
\label{fig:gc_summary}}
\end{figure}

\section{Demonstrations with Real Data}
\label{sec:real}
We now turn to demonstrations of \pro{} usage with real data.

\subsection{Globular Cluster Spectroscopy and Photometry}
\label{sec:real:ggc}

Globular clusters have long been used as a strong test of stellar population models \citep[e.g.,][]{bruzual03, koleva08}.  They are generally single-age, single-metallicity populations, making them real-world examples of simple stellar populations.  Many clusters in the Milky Way have high-quality ages based on the main sequence turnoff and metallicities determined from high-resolution spectroscopy of individual stars.  However, globular clusters present unique challenges -- at low metallicities they have very prominent blue horizontal branches (BHB), which may be a consequence of the unusual light-element abundance variations that have been detected in nearly all clusters \citep{Carretta09}.  If these unusual features are confined to the cluster population, then they may be less useful as tests of stellar population models.

Here we fit the integrated spectra and photometry of a sample of Milky Way globular clusters with \pro{}.  The sample is drawn from the spectral atlas of \citet{schiavon05}, who conducted drift-scan long-slit spectroscopy of 40 globular clusters.  These spectra span the wavelength range 3400-6400\AA{}, with a spectral resolution that is wavelength dependent but averages 3.2\AA{} FWHM.  The nominal S/N is often greater than 100.  We take \emph{UBVRI} photometry from \citet{harris96}, who provide total (integrated) magnitudes of clusters. We assign uncertainties of 10\% to the photometric fluxes.

To model this data we treat each cluster as a single age stellar population. The free parameters of this population are its mass, age, and stellar metallicity $\feh$ (assuming scaled solar abundances). We fix the distance to the value given by \citet{harris96} since for integrated SEDs at non-cosmological distances the distance and mass are perfectly degenerate. We broaden the model library spectra to match the instrumental resolution as reported by \citet{schiavon05}, and fit for an additional stellar velocity dispersion $\sigma_v$ as well as a systemic redshift. Extinction is modeled with the attenuation curve of \citet{cardelli89}, with the normalization of this curve left as a free parameter.  We also multiply the model spectrum by a maximum likelihood 15th-order Chebyshev polynomial at each likelihood call, to account for imperfections in the spectrophotometric calibration.

In Figure \ref{fig:gc_demo} we show an example of the spectroscopic and photometric data, the residual of the posterior prediction for the SED from the data, the posterior prediction for the calibration vector, and the inferred posterior probability distribution functions for the key physical parameters. The photometry is typically well fit by the models, and the spectra can be fit simultaneously when accounting for offsets from the photometry that vary smoothly with wavelength by $\sim 10$\% and inflation of the nominal noise estimates by factors $\sim 2$.

In Figure \ref{fig:gc_summary} we show a comparison of the parameters that we infer for many GCs, compared to literature values as compiled by \citet{koleva08} and to values inferred from the same spectra using the UlySS code \citep{koleva09}.  The literature values are based on stellar CMD fitting and metallicity measurements of individual stars.    We find that the inferred  metallicities from combined spectroscopy and photometry with \pro{} are in good agreement with literature values, even in the presence of un-modeled $\alpha$-element enhancements. The median offset is 0.04 dex and the r.m.s. scatter is $\sim 0.2$ dex, including outliers at very low metallicity where the stellar libraries in our models are incomplete.  Good agreement is also achieved between the inferred and literature values for the reddening ($A_{\rm V}$).

Previous work has shown the integrated-light spectral fitting of globular cluster data returns generally unreliable ages \citep{koleva08,Conroy18}.  We find similar results with \pro{}.  The primary issue is the impact of hot BHB stars which, since they are not included in the modeling, will tend to result artificially young inferred ages \citep[see ][for a discussion of this effect]{koleva09}.  Indeed, in Figure \ref{fig:gc_summary} one sees that the clusters with the most significant BHB populations (as inferred from the HBR parameter) have the youngest ages from \pro{}.  There are also some cases where integrated-light measurements (both from \pro{} and \citealt{koleva08}) result in maximally-old ages.  The origin of this shortcoming is unclear, but could be due to other model systematics such as un-modeled $\alpha$ enhancement or other limitations in the isochrones.  We note that differences between 10 and 14 Gyr are quite subtle in integrated light.

Overall, we find that \pro{} performs as well as other stellar population fitting programs when applied to globular clusters.

\begin{figure*}
\begin{center}
\includegraphics[width=0.9\textwidth]{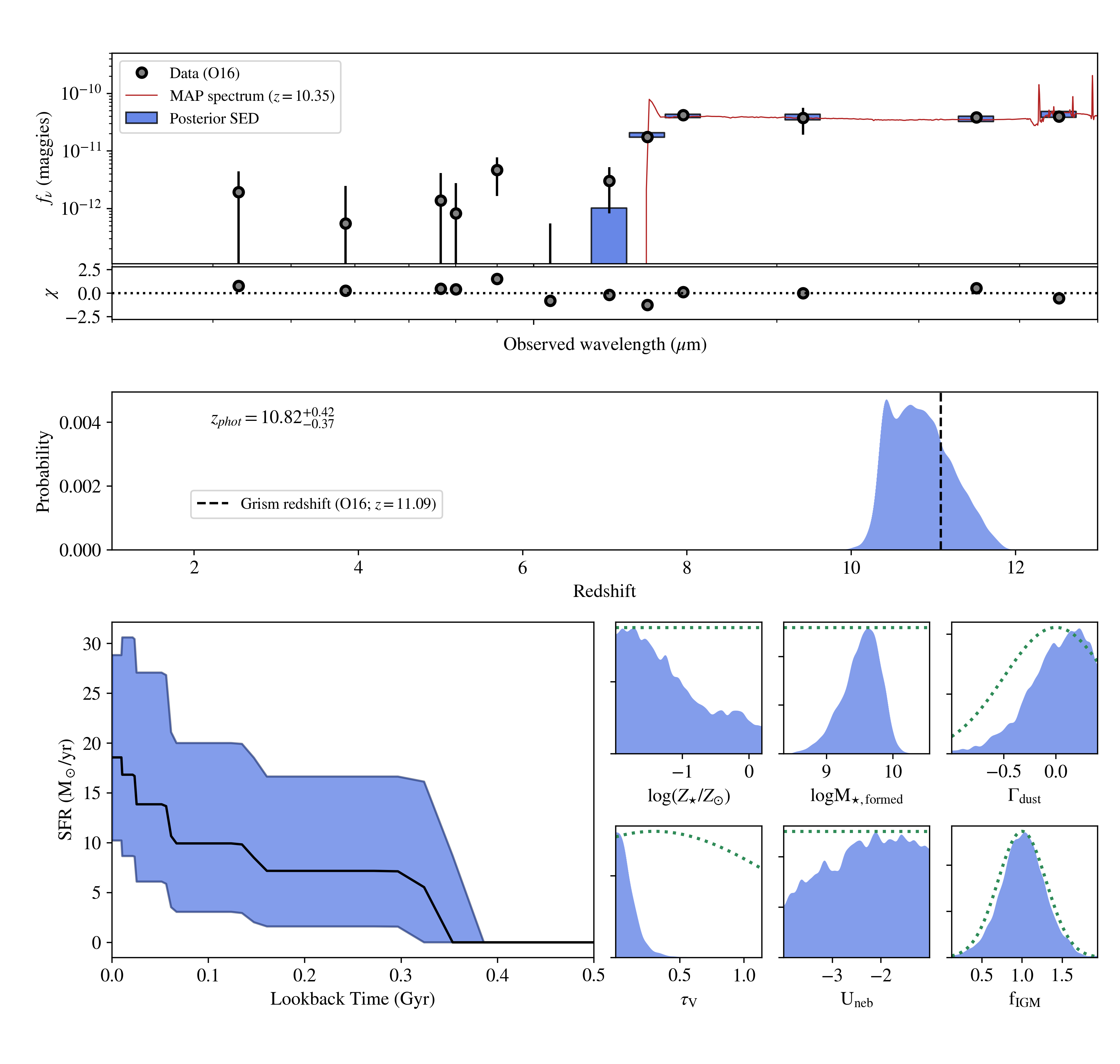}
  \caption{Photometric redshift inference with \pro{}.  It is possible to include the redshift as free parameter in the model.  Here we show the results of inference from the photometry of GNz-11.
  {\it Top row:} The fitted observed photometry (orange with black error bars) from \citet{oesch16} as well as the posterior PDFs for the true photometry (blue rectangles) and the predicted spectrum of the most probable sample from the Monte Carlo chain (red) which has a redshift of 10.35.
  {\it Second row:} Uncertainty normalized residual ($\chi$) between the observed photometry and the predicted photometry for the most probably Monte Carlo sample.
  {\it Third row:} The posterior PDF for the redshift (blue shaded region).  The prior is uniform over the range $2 < z < 13$.  Also indicated is the redshift inferred by \citep{oesch16} from HST grism data (black vertical dashed line). The text gives the median of the redshift PDF, with uncertainties derived from the 16th and 84th percentiles.
  {\it Fourth row:} Marginalized posterior PDFs for selected model parameters (blue histograms) as well as the prior probabilities for each parameter (green dotted lines)
  {\it Fifth row:} Inferred SFH.  The SFH is modeled with 5 bins in lookback time, which change in width depending on the model redshift.
\label{fig:gnz11}}
\end{center}
\end{figure*}

\subsection{Photometric Redshift: \gnz}
\label{sec:gnz11}
Inferring galaxy redshifts from photometric data is one of the core applications of stellar population models.  Here we demonstrate the use of \pro{} for this task, using the high-redshift galaxy candidate \gnz{}.  This object was selected as a high-redshift galaxy candidate on the basis of photometry with {\it Hubble Space Telescope} and {\it Spitzer} in the GOODS-North field, and followup grism spectroscopy with HST confirms a redshift $z=11.09$ \citep{oesch16}.

The model used to fit the data includes the galaxy redshift as a free parameter.  The other free parameters are the total stellar mass formed, the stellar and nebular metallicity, the nebular ionization parameter and three parameters describing the dust attenuation.  Due to the potential for the photometry to probe the rest-frame $\lambda<1216$\AA{} spectrum at high redshift we include a redshift dependent IGM attenuation following \citet{madau95}. This includes a free parameter that scales the total IGM opacity, intended to account for line-of-sight variations in the total opacity (see Appendix \ref{apx:fsps} for details). The SFH is modeled with 5 temporal bins, with the continuity prior discussed in \S\ref{sec:illustris}.  The exact width and extent of these bins are dependent on the redshift of a given model, bounded such that no stars are older than the age of the universe.  The priors are similar to those given in Table \ref{tbl:nbands_parameters}.  For the multiplicative scaling of the IGM attenuation curve we adopt a Gaussian prior distribution centered on 1, with dispersion of 0.3.  The ideal redshift prior would be the product of the redshift-dependent differential cosmological volume element and the integral of the redshift-dependent luminosity function above some limiting magnitude.  However, for clarity and due to a lack of knowledge about the luminosity function at these epochs, we adopt a uniform prior over the range $1 < z < 13$. There are 13 free parameters in the model.

The results of fitting this model to the photometric data are shown in Figure \ref{fig:gnz11}.  The photometric data are well fit, and the photometric residuals from the sample with the highest posterior probability are consistent with the stated errors.  The posterior probability distribution for the redshift shows a broad peak at $z=10.8^{+0.4}_{-0.4}$, and the highest posterior probability sample has $z=10.35$ though many samples with nearly equal probability are distributed from $z\sim10.3$-$11.4$.  This is in excellent agreement with the redshift of $z=11.09 \pm 0.1$ inferred from HST grism data. We do not find significant additional or secondary peaks in the posterior distribution of redshift using this model, though a \emph{much} weaker peak is present at $z\sim 2$. Posterior distributions for the SFH and several additional model parameters suggest a rising SFH, a formed stellar mass of $\sim 10^{9.6} \msun$, low dust attenuation, a weak constraint on the stellar metallicity, and no constraint on the IGM opacity scaling factor beyond the imposed prior distribution.

\begin{figure*}
  \begin{center}
  \includegraphics[width=0.8\textwidth]{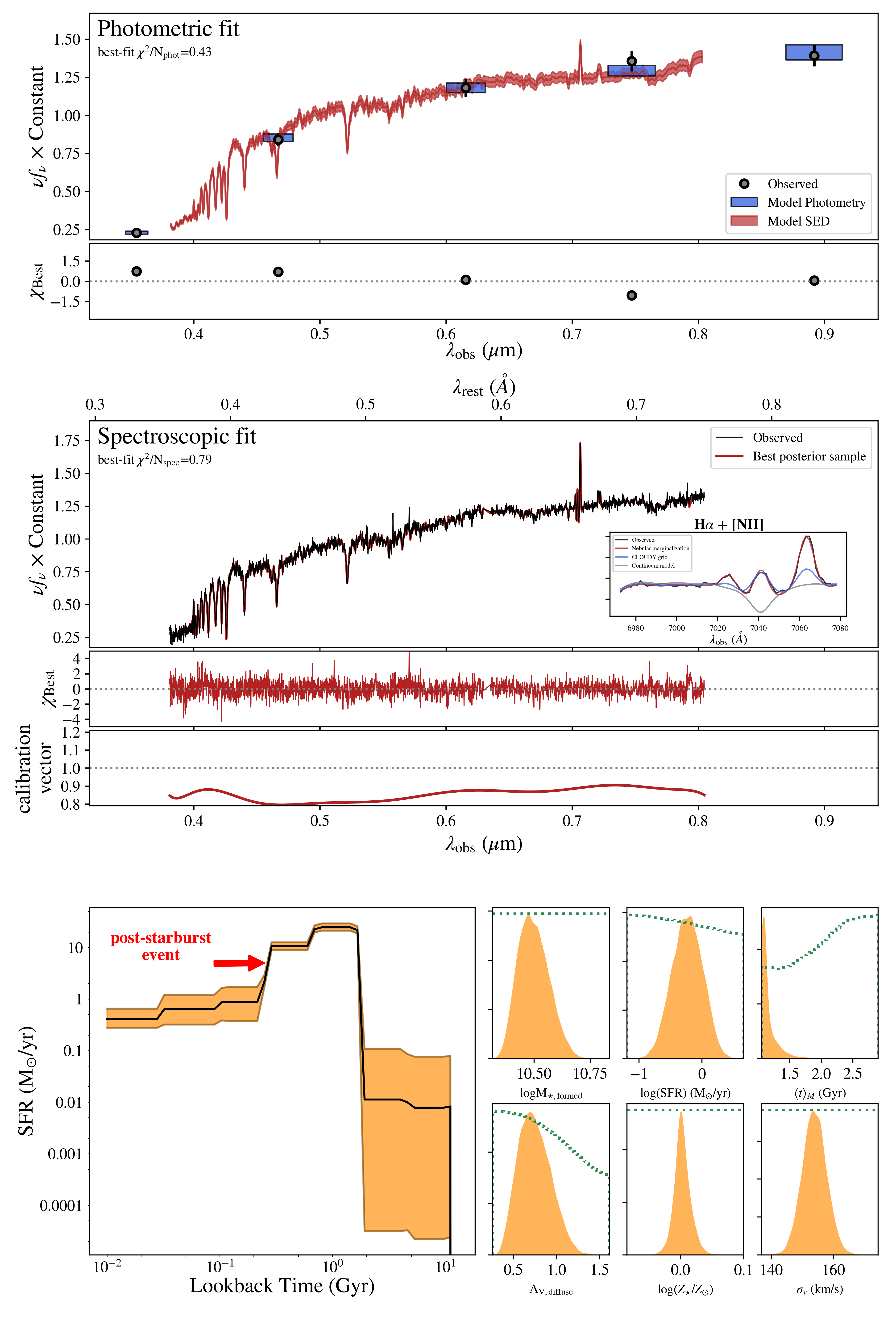}
  \caption{A simultaneous \pro{} fit to the spectroscopy and photometry of an SDSS post-starburst galaxy \citep{goto07}.
    {\it Top row:} The observed $ugriz$ photometry (black), compared to the model posterior prediction for the photometry (red) and the model spectrum of the most probable sample.
    {\it Second row:} The observed SDSS spectrum (black), compared to the model posterior prediction for the spectrum (red). The inset panel zooms in on the H$_{\alpha}$ and [NII] emission lines, showing the continuum model (grey), emission lines predicted from the CLOUDY grid (blue), and emission lines from the nebular marginalization model (red). The lower panel shows the normalization vector between the spectroscopy and photometry.
    {\it Third row:} Marginalized 1D posterior distributions for selected model parameters, along with their median and $1\sigma$ confidence intervals.
    {\it Fourth row:} The recovered star formation history. The factor-of-ten decrease in the star formation rate at $\sim$300 Myr is the post-starburst event.
    \label{fig:sdss_psb_pdf}}
    \end{center}
\end{figure*}

\subsection{SDSS Post-Starburst Spectrum}
\label{sec:sdss}
We fit a post-starburst galaxy from the SDSS IV \citep{gunn06,blanton17}, chosen to highlight \pro{}'s capability to perform a simultaneous fit to spectra and photometry. A post-starburst galaxy is characterized by strong hydrogen absorption lines, indicating it has decreased its star formation activity sharply in the past $\sim500$ Myr. We  select a post-starburst galaxy to demonstrate both the flexible SFH model and the flexible emission line modeling available within \pro{}.

We fit SDSS J133751.29+354755.7 from the post-starburst catalog of \citet{goto07}, specifically targeted for its high signal-to-noise spectrum and substantive emission line infilling. We fit the $ugriz$ photometry and optical spectrum \citep{smee13} from SDSS Data Release 16 \citep{ahumada20}. A noise floor of 5\% is enforced on the photometry consistent with  uncertainties in the underlying stellar models.

A 19-parameter physical model is used. The total mass, stellar metallicity, and velocity dispersion are set free. The redshift is allowed to vary within 0.01 around the SDSS catalog value of $z=0.073$.  The dust model is the two-parameter \citet{charlot00} model, plus a flexible attenuation curve \citep{noll09} with the UV bump tied to the slope of the curve \citep{kriek13}. An eight-bin non-parametric SFH is used, with the time bins spaced logarithmically and a continuity prior applied \citep{leja19a}. Nebular emission is implemented using the \citet{byler17} grid, with metallicity, ionization parameter, and nebular velocity dispersion set free. Nebular line marginalization is also turned on, as described in Appendix \ref{apx:nebular}; in brief, the amplitude of each emission line is marginalized at each model call to fit the data. The prior on the emission line fluxes is taken as a Gaussian, with both the center and the standard deviation taken to be the expected line luminosities from the \citet{byler17} physical model.  A polynomial vector is used to normalize out the continuum during each model call, following Section \ref{sec:ingredients:cal}. Finally, the pixel outlier model and jitter term from Appendix \ref{apx:noisemodel} are both adopted.

The results of this fit are shown in Figure \ref{fig:sdss_psb_pdf}. \pro{} is able to fit the data with high precision; the best-fit model produces excellent $\chi^2$ results for both the photometry and the spectrum. Though they are included in the fit, neither the spectral noise inflation nor the pixel outlier model are assigned any significant weight. As expected, a significant post-starburst event is detected within the previous $\sim$300 Myr, suggesting a factor-of-twenty decrease in the ongoing rate of star formation. The mass-weighted age is 1.11 Gyr, indicating a relatively short formation phase for this galaxy, and the dust attenuation is significant at A$_V = 0.74$.

The spectral fit includes a zoom-in on the H$\alpha$ and [N II] doublet. The [NII]/H$\alpha$ ratio is very high, as is typical of post-starburst objects \citep{yan06,yang06,french15}. This is consistent with SDSS J133751.29+354755.7 being a LINER; in such objects, the nebular emission lines are thought to be partially or fully powered by evolved stars, AGN, or shocks (e.g., \citealt{kewley06}). This is in contrast to the built-in assumption in \fsps{} that emission lines are powered by young stars only. This difference has the potential to cause two distinct issues: first, additional young stars would have to be added in order to fully power this emission, and second, the observed [NII]/H$\alpha$ ratio is challenging and likely impossible to produce with star formation alone. Indeed, the prediction from the CLOUDY grid is indicated in blue, and is a poor fit to the data. The nebular marginalization technique, indicated in red, solves both of these issues by allowing flexible extensions from the standard CLOUDY grid. It is suggested to allow nebular marginalization in spectral fits whenever there is a chance that the emission lines are powered by sources that do not lie on the standard CLOUDY grid.

\section{Discussion}
\label{sec:discussion}

\subsection{Lessons Learned and Practical Advice}

The development of \pro{} began in 2014, and it has undergone continual development, refinement, and refactoring since.  This development is usually guided by areas of practical focus, as the diversity and complexity of questions and data to which SED fitting is applied makes general solutions nearly impossible.  Because of this, \pro{} has tended to become \emph{more} modular with time, in order to maintain flexibility while adding new capabilities.  A benefit of this modularity is that changing or extending the code to tackle new and more detailed data is relatively straightforward. In a sense, \pro{} is a toolbox for SED fitting.
The tradeoff for flexibility is often computational speed; some tasks that for particular problems could be optimized with a different code structure are instead less efficiently implemented in a modular structure.

Priors are an essential ingredient in SED fitting.
The SEDs of single age stellar populations change subtly with both age and metallicity, making SFH inference from an integrated SED particularly difficult.  The SEDs clearly contain information; the question is how to extract the available information in the face of substantial degeneracies.
The solution is to impose priors on the SFH, either explicitly or through a particular SFH parameterization, that minimize the impact of these degeneracies where the data are not informative, yet still allow the data to ``speak".  Nevertheless, it is important to map the remaining degeneracies and to consider the effect of the adopted prior.

Priors on dust emission parameters can also play an important role when the far-IR SED is incompletely sampled.  The inferred total infrared luminosity depends sensitively on these parameters in this case, and through the imposed energy balance between dust emission and dust attenuation can affect many other parameters (e.g., Appendix C of \citet{leja17}).  It is important to adopt priors that reflect prior knowledge about the distribution of these parameters, for example derived from subsamples with more complete IR coverage. Through the Bayesian framework this prior knowledge can be propagated through to the posterior PDFs.

While many of the uncertainties and degeneracies that can be explored through a Bayesian framework are important to consider, significant \emph{systematic} uncertainties in the stellar population synthesis models remain. The modularity of \fsps{} grants the ability to explore some of these as model parameters, for example the temperature distribution of horizontal branch stars or the luminosity of the TP-AGB phase. However, many other uncertainties are ``baked in'' to the stellar populations codes: these include binary evolution, stellar rotation, ionizing fluxes and more generally the blue/UV spectra of low metallicity populations, other isochrone details, and the detailed physics of nebular emission. Results that depend on the details of these calculations should be treated with caution.

\subsection{Future directions}
\label{future}

As described above, \pro{} is under continual development, and we expect to add new capabilities in the future.  Of special emphasis is computational speed.  The computational time required is set by the product of two factors: the speed at which stellar populations models are constructed, and the number of models required to adequately sample the posterior distribution (which increases with the dimensionality of the model). In the former category, we are exploring the use of emulators \citep[e.g.][]{alsing20} to produce models quickly, with the disadvantage that the emulator must be trained on pre-computed (and therefore largely inflexible) grid.  In the latter category, we are considering new Monte Carlo techniques based on gradients.  Such techniques will also allow larger model dimensionality.

While the underlying \fsps{} stellar population synthesis models used by \pro{} include most physical components of a galaxy -- stellar emission, nebular emission due to stellar photoionization, dust attenuation, dust emission, and the emission of AGN dust torii -- additional components may be added.
These include ISM absorption, nebular emission from shocked ISM gas, free-free emission from HII regions, radio synchrotron emission, and X-ray through optical emission from AGN.  Additionally, Lyman-$\alpha$ radiative transfer models may prove important for modeling high redshift galaxy spectra in detail. For some of these components straightforward ab-initio physical models exist.  For others, empirical relationships may have to be adopted.

We will also pursue improved treatment of the existing components.  For the stellar populations, this includes the addition of $\alpha$-element enhancement to the underlying SPS models, in both the isochrones and the spectra.  We also plan to expand the capability to model age dependent stellar metallicity and dust attenuation, as well as additional SFH forms. Finally, more flexible and self-consistent nebular emission models may prove useful.

Performing optimized SFH inference is a perennial challenge. The SFH ($=\mathrm{SFR}(t, Z)$) is fundamentally a density; inference of densities from samples or integrals thereof is a notoriously difficult, and perhaps even ill-posed, problem. A variety of solutions to the problem have been proposed.  These can all be considered various choices about the priors that are placed on the form or properties of the SFH.
Ideally we would combine a prior with a very large number of time bins, or anchor points, without consideration of the uniqueness of the spectra for each bin. Then, the data would be able to dictate which bins (if any) were well constrained while the prior would take over where the data do not provide information.  However, sampling in such high-dimensional spaces is difficult, though solutions based on scalable gradient-based sampling approaches (Hamiltonian Monte Carlo) are under development.  Additionally, more complex priors on the \emph{shape} of the SFH -- for example corresponding to the correlation function between SFRs in different time periods \citep[e.g.][]{kelson16, tacchella20} may be implemented.

Multiple spectra described by the same physical parameters but substantially different instrumental parameters (e.g. spectral resolution or calibration) can not yet be fit simultaneously in \pro{}, though this capability will be added in the future.  This can occur due to multiple exposures of the same object taken under different observing conditions, or to spectra from different instruments or spectroscopic orders, or both.

In summary, as the available data increase in quality and diversity, SED models and inference techniques have had to improve as well, while growing in complexity. Significant progress has been made, leading to the ability to model and make inferences from many different kinds and combinations of galaxy observations. Efforts are underway to tackle remaining challenges. These efforts are necessary to fully exploit the information content of rich datasets that will be provided by future facilities, such as \emph{James Webb Space Telescope}.

\acknowledgements
We are grateful for discussions early in the project in with Dan Weisz, Dan Foreman-Mackey, and David Hogg, as well as conversations with Adam Carnall and Sandro Tacchella.  We are also grateful for early testing of the code and suggestions by John Moustakas, Antara Basu-Zych, Dylan Nelson, Imad Pasha, Tom Zick, Song Huang, and Johnny Greco. BDJ and CC acknowledge support from the Packard Foundation and NSF grants AST-1313280 and AST-1524161. This research made extensive use of NASA's Astrophysics Data System Bibliographic Services. Computations in this paper were run on the FASRC Cannon cluster supported by the FAS Division of Science Research Computing Group at Harvard University.

Funding for the Sloan Digital Sky Survey IV has been provided by the Alfred P. Sloan Foundation, the U.S. Department of Energy Office of Science, and the Participating Institutions. SDSS acknowledges support and resources from the Center for High-Performance Computing at the University of Utah. The SDSS web site is www.sdss.org.

\software{dynesty \citep[v1.0.0][]{speagle20},
          emcee \citep[v2.2.1][]{foreman-mackey13},
          FSPS \citep[v3.1][]{conroy09,conroy10a},
          python-FSPS \citep[v0.3.0][]{pyfsps},
          Astropy \citep[v4.0][]{astropy1, astropy2},
          NumPy \citep[v1.17.2][]{numpy},
          SciPy \citep[v1.3.1][]{scipy},
          matplotlib \citep[v3.1.1][]{matplotlib},
          IPython \citep{ipython},
          f2py \citep{f2py}
         }


\begin{appendix}

\section{\fsps{} Updates}
\label{apx:fsps}

There have been many updates and improvements to the \fsps{} code since 2010.  In this section we provide a brief overview of the main features now available in the code.

Isochrone libraries are available from the MIST \citep{choi16}, PARSEC \citep{Bressan12}, BaSTI \citep{Cordier07}, Padova \citep{marigo08}, Geneva \citep{Meynet94}, and BPASS \citep{eldridge17} models.  The BPASS models are provided as simple stellar populations and are therefore treated differently than the other isochrone sets: for example, the isochrone-based parameters that affect the strength of the TP-AGB cannot be applied to the BPASS models, nor can the IMF be changed.

As in \citet{conroy10b}, the available base spectral libraries include BaSeL and MILES.  Those libraries are supplemented with the TP-AGB library from \citet{Lancon00a}, the post-AGB library from \citet{Rauch03}, and the WMBASIC hot star library from \citet{eldridge17}.

There are a variety of options for the dust absorption model including the Milky Way extinction curve of \citet{cardelli89} with variable slope and UV bump strength, empirical attenuation curves from \citet{Calzetti00} and \citet{kriek13}, the two-component dust model of \citet{charlot00} with separate normalization of each component and adjustable attenuation curve slopes, and the radiative transfer models from \citet{Witt00}.  ``Diffuse" dust emission is provided by the \citet{draine07} templates.  Circumstellar dust around AGB stars is implemented via the \citet{Villaume15} templates, while dust associated with AGN is modeled using a simplified implementation of the \citet{nenkova08b} templates.

Nebular emission associated with HII regions is included using Cloudy lookup tables computed by \citet{byler17}.  Attenuation by the intergalactic medium (IGM) follows the prescription described in \citet{madau95}, with an optional overall scaling factor to account for line-of-sight variations.

\section{SFH Integrals and Parametric Forms}
\label{apx:fsps:sfh}

The fundamental equation for computing the spectra of composite stellar
populations is
\begin{eqnarray}
F_\lambda & = & \int_0^{t_{univ}} dt \int_{Z_{min}}^{Z_{max}} dZ \, \psi(t, Z) \, s_\lambda(t, Z) \, e^{-\tau_\lambda(t, Z)}
\end{eqnarray}
where $t$ is the \emph{lookback time}, $\psi(t, Z)$ is the SFH, $s_\lambda(t, Z)$ is the spectrum of a unit mass of stars of age $t$ and metallicity $Z$, and $\tau_\lambda(t, Z)$ is the \emph{effective} dust opacity towards stars of age $t$ and metallicity $Z$.

In practice, this is approximated by a sum over discrete \emph{simple stellar populations} (SSPs). These SSPs are the $s_\lambda(t, Z)$ for $J$ specific values of $t$ (indexed as $t_j$) and $K$ specific values $Z$ (indexed as $Z_k$). They are calculated using isochrones and stellar spectral libraries. The sum is then
\begin{eqnarray}
F_\lambda & = & \sum_{j=1}^J \, \sum_{k=1}^{K} \, m_{j,k} \, s_{\lambda, j,k} \, e^{-\tau_{\lambda, j,k}} \quad ,\\
s_{\lambda, j,k} & = & s_\lambda(t_j, Z_k) \quad ,
\end{eqnarray}
where the $m_{j,k}$ are the formed masses or \emph{weights} of each SSP.

It is tempting to use trapezoidal integration of the SFR at $t_j, Z_k$ to calculate these weights. However, if the SFR has a discontinuous derivative or is highly-nonlinear between the SSP points, this can lead to substantial errors. Increasing the effective temporal resolution of the SSPs via interpolation can mitigate (but not solve) this problem, at the expense of the increased memory needed to store the interpolated SSPs, and the increased computational time needed to sum them with the appropriate weights.

A more robust and scalable solution is to integrate the product of the interpolation weights and the SFH. That is, if a spectrum at time $t_j < t < t_{j+1}$ can be written as
\begin{eqnarray}
s_\lambda(t) & = & a_j(t) \, s_{\lambda}(t_j) + b_{j+1}(t) \, s_{\lambda}(t_{j+1}) \quad , \\
a_j(t) & = & \frac{\log (t_{j+1}) - \log(t)}{\log(t_{j+1}) - \log(t_j)} \quad , \nonumber \\
b_{j+1}(t)  & = & \frac{\log t - \log(t_j)}{\log(t_{j+1}) - \log(t_j)} \quad , \nonumber
\end{eqnarray}
where we have chosen a linear interpolation in $\log t$ (because changes in the SSP fluxes are generally linear in $\log t$, not in linear time), then we wish to calculate
\begin{eqnarray}
m_j & = & \int_{t_j}^{t_{j+1}} dt \, \Psi(t) \, a_j(t)  + \int_{t_{j-1}}^{t_{j}} dt \, \Psi(t) \, b_j(t) \, .
\end{eqnarray}
This gives the contribution to the weights of the SFR in the bins that are older and younger than $t_j$, and for parametric SFHs this can be solved analytically for each $j$. This is what is used in the \fsps{} code for the available parametric SFHs, as of v3.0. Combinations of parameterized SFHs can be easily considered by calculating the above quantity for each component and summing the $m_j$ values with the appropriate scaling to retain the proper mass fractions in each component. Similarly, for tabular SFHs we make the assumption that the SFR is a linear function between the tabulated time points, and solve the equation above for each $t_j$ and segment separately, accumulating the weights $m_j$ from each segment.

\section{Smoothing}
\label{apx:smoothing}

When comparing model galaxy spectra to observed spectra there are (at least)
three spectral smoothing resolutions (or smoothing kernels) that must be considered.  These are:
	1) the line-of-sight velocity distribution (LOSVD) of stars in the restframe of the galaxy (${\rm L}_g$);
    2) the observed frame line-spread function of the instrument (${\rm L}_i$); and
    3) the restframe line-spread function of the model spectral library  (${\rm L}_\ell$)
.
Along with any imposed smoothing of the model spectrum, these combine to create
the line-spread functions of the data (${\rm L}_{d}$) and the model (${\rm L}_{m}$) as follows:
\begin{eqnarray}
{\rm L}_{d}(\lambda_o) & = & {\rm L}_{i}(\lambda_o) \conv {\rm L}_{g}(\lambda_r) \quad , \nonumber \\
{\rm L}_{m}(\lambda_o) & = & {\rm L}_{\ell}(\lambda_r) \conv {\rm L}_{k}(\lambda_o) \quad ,\nonumber \\
\lambda_o & = & (1+z) \, \lambda_r \quad ,
\end{eqnarray}
where $\lambda_o$ is the observed frame wavelength,
$\lambda_r$ is the restframe wavelength,
$L_{k}$ is the kernel corresponding to extra smoothing applied while modeling the data,
and $\conv$ denotes convolution.
While the LOSVD of the galaxy can be handled by straightforward and fast smoothing in
velocity space \citep[e.g.][]{cappellari17}, the instrument and library resolution are both potentially complicated functions of observed-frame and rest-frame wavelength respectively.
This can be difficult to implement in a speedy convolution algorithm.

When ${\rm L}_{i} = {\rm L}_\ell$ or both are narrow compared to ${\rm L}_g$ they can be neglected and the best match between the data and model is obtained when ${\rm L}_{k} \sim {\rm L}_g$. Varying ${\rm L}_k$ to determine the best match can be done using fast FFT based smoothing in velocity space. However, when the LOSVD of the galaxy is comparable to or only slightly broader than the instrumental or library LSF (at any observed wavelength) then these latter effects must be taken into account.  The possible wavelength dependence of the instrumental and library resolutions makes this difficult, as standard FFT-based algorithms do not admit smoothing kernels that are wavelength dependent.  The usual solution is to use slower brute-force convolution algorithms to smooth the library spectra \emph{once}, before any other calculations are made or any fitting is done, such that ${\rm L}^\prime_\ell(\lambda_r) \sim {\rm L}_i(\lambda_o)$. This is possible provided the library everywhere has higher resolution than the instrument, that ${\rm L}_i$ is well known, and that the redshift of the model does not vary significantly.

If any of these conditions are not met -- for example if one wishes to infer and marginalize over uncertainties in the instrumental LSF -- then another approach must be used. A solution is to include, in each generation of the model, a convolution by the kernel ${\rm L}_{k}$ defined by the difference between the rest-frame library resolution and the observed frame instrumental and physical resolution.  This potentially wavelength dependent difference kernel may depend on the modeled systemic velocity.  Because the convolution is done at each model generation step, this scheme also allows for simultaneous fitting of the LSF.

In \pro{} we allow for both options: either smoothing the library once, before fitting, by the difference between the observed-frame instrumental and rest-frame library resolution or smoothing at each model generation step by this difference, calculated on the fly for the model systemic velocity, velocity dispersion, and instrumental LSF parameters.  We assume that both the instrumental and library resolution can be approximated by Gaussians at each wavelength, so that the difference kernel is also represented by a wavelength dependent Gaussian.

It is desirable to have algorithms that can quickly and accurately produce spectra smoothed by a wavelength dependent kernel, so that instrumental smoothing can be done \emph{during} the fitting process in a flexible way. Since the difficulty with FFT-based algorithms in this case is that the kernel is varying as a function of wavelength, the solution is to resample the model spectrum onto a space in which the kernel is \emph{not} varying (analogous to resampling a spectrum from a regular wavelength grid to a regular velocity grid when smoothing by a LOSVD.)

Assuming the line-spread function can be described by a Gaussian, this can be accomplished by defining a coordinate system from the cumulative distribution function of the inverse of the dispersion as a function of wavelength
\begin{eqnarray}
x(\lambda) & \equiv & K \, \int_{\lambda_{min}}^{\lambda} \frac{d\lambda^\prime}{\sigma(\lambda^\prime)} \quad ,
\end{eqnarray}
where $\lambda$ is the wavelength, $\sigma(\lambda)$ is the width of the Gaussian LSF at any wavelength, and K is a normalization constant. A regular sampling in $x$ will provide an equal number of samples per $\sigma$, even as $\sigma$ changes with wavelength.  Thus, convolving a spectrum resampled onto an even grid in $x$ with a single Gaussian (in $x$) will produce a spectrum with a smoothing in $\lambda$ that is $\lambda$ dependent.

The integral that defines $x$ can be done analytically or, for more flexibility, numerically.  The constant $K$ and the final spacing in $x$ should be chosen so that the interpolated spectrum adequately samples at least the output line-spread function, and preferably the input LSF, for some definition of adequate.
Such an algorithm is implemented within \pro{}. Further improvements to the smoothing functionality in \pro{} may include a larger variety of smoothing kernels (including higher order LOSVD moments and kernels appropriate for pixelization.)

\section{Alternative Noise Models}
\label{apx:noisemodel}

In addition to the default assumption of known Gaussian independent noise described in \S\ref{sec:likelihood}, we have also implemented more complex likelihood calculations.  These allow for properties of the noise itself to be inferred, or for the presence of correlated noise to be modeled.  This is accomplished though a flexible noise covariance matrix construction.  The $\ln$-likelihood for the more complex noise model is
\begin{eqnarray}\displaystyle
\ln \like(d \given \Theta, \gamma) &=&
-\frac{1}{2}\, \left[ \transpose{\Delta} \, C^{-1}(\gamma) \, \Delta + \ln \det{2\pi\, C(\gamma)} \right] \label{eqn:complex_likelihood}\\
\Delta &=&  d - \mu(\Theta)
\quad ,
\end{eqnarray}
where $d$ is the data vector of length $n$, $\mu$ is the mean physical and instrumental model which depends on the parameters $\Theta$, $\Delta$ is the residual, and $C$ is the variance-covariance matrix which depends on the noise model parameters $\gamma$.

The parameters $\Theta$ generally include the physical parameters $\theta$. They may also include instrumental effects such as wavelength-scale distortions or the instrumental resolution as parameters $\phi$.

In \pro{} the covariance matrix is specified as a sum of weighted kernels applied to given \emph{metrics}, where the kernels, weight vectors, and metrics can be specified at run time.  More precisely, we use the following formulation of $C$
\begin{eqnarray}\displaystyle
C_{i,j} &=& \sum_\ell w_{i,\ell} \, w_{j,\ell} \, K_\ell (x_i, x_j, \gamma_\ell)
\quad ,
\end{eqnarray}
where $w$ is the weight vector of length $n$, $x$ is the metric (also of length $n$), $K_\ell$ is the $\ell^{th}$ kernel, and $\gamma_\ell$ are the noise parameters for that kernel.

The kernels $K$ can in principle be anything that results in a positive semi-definite matrix when multiplied by the weight vectors $w$. The following are available by default in \pro{}:
\begin{itemize}
\item \emph{uncorrelated:} $K(x_i, x_j \given a) = a \, \delta_{i,j}$.
This is a simple diagonal matrix parameterized by an amplitude $a$ for the diagonal elements.  All other elements are 0.

\item \emph{photo-cal:} $K(x_i, x_j \given a) = a$ for $i,j$ corresponding to user-specified photometric bands. This adds correlated noise to these bands with amplitude $a$. All other matrix elements are 0.

\item \emph{exp-squared:} $K(x_i, x_j \given a, L) = a \, e ^{-(x_i-x_j)^2/(2L^2)}$.
This is the exponential squared kernel, parameterized by an amplitude $a$ and scale length $L$.

\end{itemize}

A simple use for this more complicated noise model is to incorporate uncertainty in the noise estimates themselves. This is accomplished by allowing the matrix $C$ to depend on parameters $\gamma$ that are then inferred and can be marginalized over. For example, to allow for the possibility that the supplied uncertainties are underestimated by a constant but unknown factor $b$ a diagonal kernel $K_1(x_i, x_j) = b^2 \delta_{i,j}$ can be combined with weights $w_{i,1}$ given by the nominal supplied uncertainties, where $\delta_{i,j}$ is the Kronecker-$\delta$.  Increasing $b$ will decrease the first term in the likelihood of Equation \ref{eqn:complex_likelihood}, but this is eventually balanced by an increase in the normalization given by the determinant of $C$, preventing the uncertainties from becoming arbitrarily high. Inferring such an inflation in the uncertainties can be useful for providing more accurate final parameter uncertainties.  An additive noise component, i.e. a noise floor, can also be incorporated with an additional diagonal kernel $K_2(x_i, x_j) = a \delta_{i, j}$ and weights $w_{i, 2} = 1$.

The covariance matrix can be used to account for correlated noise in the data. Correlated noise can be induced during extraction of 2-D spectra, smoothing of the data, or interpolation to a particular wavelength grid \citep[e.g.][]{bolton10}.  When the pixel covariances are known from error propagation during extraction, they can be supplied in matrix form as the kernel $K$, with weights $w_i=1$.

Sky subtraction residuals are likely to be correlated over at least several pixels and may also be treated as additive covariant noise \citep[e.g.][]{carnall19b}.  The scale length and amplitude of these correlations may be supplied or inferred.

Finally, the use of weight vectors also allows calibration to be modeled as covariant or correlated noise.  Here the covariant noise may be due calibration uncertainties or to inaccuracies in the spectral models. Since calibration is multiplicative with respect to flux, modeling calibration noise can be accomplished by using the predicted model flux vector $\mu(\Theta$) as the weight vector $w$.  This approach was used for photometry, where calibration uncertainties from different instruments or observatories are often correlated, by \citet{gordon14}.  It can be extended to spectroscopic data.  This approach has the benefit of being more flexible than than polynomials or other parameterized functions, which will introduce systematic effects if they are not a good representation of the actual spectrophotometric calibration function.  It also has the benefit that a model where the data can be described well purely by parameters unrelated to the calibration (e.g. absorption lines) will be higher likelihood than one where correlated noise is necessary. This mitigates the problem that physical information in the data may be lost to the spectrophotometric calibration model if the latter is overly flexible. However, this approach requires a realistic model for the correlations induced by mis-calibration, which can be complex, multi-scale and must be separated from correlations induced solely by data processing (e.g. by interpolation or spectral extraction.)  This approach may also incur high computational cost.

Finally, we also include a rudimentary outlier model in \pro{}. This model approximately accounts for the presence of data which deviates from the model much more than should be allowed by the noise, due for example to cosmic rays or strong, un-masked sky subtraction residuals. The outliers are modeled as a small fraction of data points that are drawn from a normal distribution centered on the model fluxes but with much larger than the nominal supplied uncertainties.  By analytically marginalizing over the exact location of the outliers \citep[e.g.][]{hogg10}, this can be expressed as
\begin{eqnarray}
\ln p^\prime(d \given \Theta, \beta) &=& (1 - \fout) \ln p(d \given \Theta, \beta) \\
& & \quad  + \, \fout \, \sum_i \frac{d_i - \mu_i(\Theta)}{2\, (s_{\rm out}\, \sigma_i)^2}
 \quad , \nonumber
\end{eqnarray}
where $\fout$ is the fraction of outlier data points, and $s_{\rm out}$ is the inflation of the nominal noise for outlier points, which defaults to 50. Both of these parameters may be supplied or inferred. At the present time this outlier model cannot be used with correlated noise models.

\section{Emission Line Marginalization}
\label{apx:nebular}
As described in \S\ref{apx:fsps} the emission line model included with \fsps{} is based on \code{Cloudy} \citep{ferland13, byler17}.  This treatment is sufficient in many contexts, but deviations of emission line ratios and luminosities from the two-parameter model implemented there may occur for a variety of reasons.  These include the presence of AGN, shocks, or other non-stellar ionizing sources, abundance variations, and complicated geometries.  Furthermore, the kinematics of the ionized gas may be different than that of the stars. It is therefore desirable to include some additional flexibility in the emission line modeling.

This is a challenge; there are 128 emission lines included in \fsps{}, and it is not practical to sample directly for the amplitude of each line. With some reasonable assumptions, it is possible to introduce analytic Gaussian emission line modeling into the \pro\ likelihood which do not require explicitly sampling for emission line amplitudes. When the prior on the amplitudes is also Gaussian, the posterior probability for the amplitude is also Gaussian and can be analytically marginalized in each likelihood call. The emission line width and central wavelength offset is sampled for as well; the offset and width are assumed to be the same for each emission line to ease the sampling, though this assumption can be relaxed by the user if necessary.

For clarity we define here a Gaussian distribution $\norm$ of the variable $x$ with mean $\mu$ and covariance matrix $\Sigma$ as
\begin{eqnarray}
\norm(\boldsymbol{x} \given \boldsymbol{\mu}, \boldsymbol{\Sigma}) & = & \frac{1}{\sqrt{\det{2\,\pi\,\boldsymbol{\Sigma}}}} \, e^{-\frac{1}{2}\transpose{(\boldsymbol{x}-\boldsymbol{\mu})} \, \Sigma^{-1} \, (\boldsymbol{x}-\boldsymbol{\mu})} \, .
\end{eqnarray}

\subsection{Analytically Marginalizing over emission lines}

First, we write the likelihood of the data, conditioned on the parameters $\theta$, marginalized over several emission line amplitudes $\alpha_j$, and including the prior on these amplitudes, as
\begin{eqnarray}
\like(D \given \theta) & = & \int \, d\ba \, \like(D \given \theta, \ba) \, p(\ba) \label{eqn:marg_like}
\end{eqnarray}
where $D$ is the data,
$\theta$ are the parameters of the model \emph{not} including the line amplitudes,
and $\ba$ is the vector of all line amplitudes $\{\alpha_j\}$.
We can then write the log-likelihood, conditional on both $\theta$ and $\ba$, as
\begin{eqnarray}
\ln \like (D \given \theta, \ba) & = & \transpose{(\bdelt - \bg\, \ba)} \, \Sigma^{-1} \, (\bdelt - \bg \, \ba) \, ,\\
\bdelt & =  & \{D_\lambda - M_\lambda\}  \nonumber  \quad ,\\
\bg & = & \{G_\lambda, j\} \nonumber \quad ,\\
G_{\lambda, j} & \equiv & \norm(\lambda \given (1 + z) \, \lambda_{0,j}, \frac{\sigma_v}{c} \, (1 + z) \, \lambda_{0,j}) \nonumber
\end{eqnarray}
where $\Sigma$ is the variance-covariance matrix of the data,
$M_\lambda$ is the model prediction without emission lines,
$\Delta$ is a vector of residuals from the line-free model,
$G$ is a normalized Gaussian profile with a given width $\sigma_v$ and center (specified as part of $\theta$),

The key is that by using $\alpha_j$ \emph{only} for the amplitude of the lines (i.e. not for the shape or center), the conditional log-likelihood is linear with respect to $\ba$.
The likelihood can then be factored into two components.
The first component is the likelihood of the model conditional on the \emph{maximum likelihood estimate} (MLE) of the line amplitudes, $\hat{\ba}$.
This scales the second component, which is a multidimensional Gaussian describing the dependence of the likelihood on the line amplitudes $\ba$.  This Gaussian is centered on $\hat{\ba}$ and has a width described by the covariance matrix $\Sigma_{\hat{\ba}}$
Our expression for the likelihood function is then
\begin{eqnarray}
\label{eqn:like_factor_vec}
\like (D \given \theta, \ba) & = & \like(D \given \theta, \hat{\ba}) \, \frac{\norm(\ba \given \hat{\ba}, \Sigma_{\hat{\ba}})}{\norm(\hat{\ba} \given \hat{\ba}, \Sigma_{\hat{\ba}})} \, .
\end{eqnarray}
The denominator is a non-unity value required to make the equality hold for $\ba=\hat{\ba}$.  When the prior is uniform ($p(\ba)$ is constant) the integral in the marginalized likelihood of equation \ref{eqn:marg_like} -- ignoring a constant factor that is independent of $\theta$ and $\ba$ -- reduces to
\begin{eqnarray}
\like(D \given \theta) & = & \frac{\like(D \given \theta, \hat{\ba})}{\norm(\hat{\ba} \given \hat{\ba}, \Sigma_{\hat{\ba}})} \quad .
\end{eqnarray}

We now derive the MLE for $\ba$. By taking the derivative of the likelihood with respect to $\ba$ and setting the result to 0, we find the MLE value $\hat{\ba}$ as
\begin{equation}
    \hat{\ba} = (\transpose{\bg} \Sigma^{-1} \, \bg)^{-1} \, \transpose{\bg} \, \Sigma^{-1} \, \bdelt \quad .
\end{equation}
We are also interested in the widths or uncertainty on the MLE value, described by the covariance matrix $\Sigma_{\hat{\ba}}$. This is obtained from the inverse second derivative of the likelihood with respect to $\ba$, resulting in
\begin{equation}
    \Sigma_{\hat{\ba}} = (\transpose{\bg} \Sigma^{-1} \, \bg)^{-1} \quad .
\end{equation}
Notably, by doing this calculation in vector space, we explicitly model possible covariances between the line amplitudes; this is important in cases where the emission lines overlap.

Using the MLE solutions for the emission line amplitudes is correct for a uniform prior on $\ba$. In some cases, an informative prior on $\ba$ is desirable. This is discussed in the next section.

\subsection{Gaussian priors on the emission line amplitudes}
To incorporate priors we can multiply the factorized likelihood in equation \ref{eqn:like_factor_vec} by the prior.  This is straightforward if the prior is another Gaussian with mean $\breve{\ba}$ and covariance  $\Sigma_{\breve{\ba}}$: that is, $p(\ba) = \norm(\ba \given \breve{\ba}, \Sigma_{\breve{\ba}})$.  The product of two Gaussians is another Gaussian; using well known formulae for the products of normal distributions we obtain
\begin{eqnarray}
\like(D \given \theta, \ba)\, p(\ba) & = & \like(D \given \theta, \hat{\ba})  \, \frac{\norm(\hat{\ba} \given \breve{\ba}, \Sigma_{\hat{\ba}} + \Sigma_{\breve{\ba}})}{\norm(\hat{\ba} \given \hat{\ba}, \Sigma_{\hat{\ba}})}\norm(\ba \given \bar{\ba}, \Sigma_{\bar{\ba}})
\end{eqnarray}
where $\bar{\ba}$ and $\Sigma_{\bar{\ba}}$ are the mean and covariance of the new Gaussian.  In fact this new Gaussian does not need to be evaluated, since the integral in the marginalized likelihood of equation \ref{eqn:marg_like} reduces to
\begin{eqnarray}
\like(D \given \theta) & = & \like(D \given \theta, \hat{\ba})  \, \frac{\norm(\hat{\ba} \given \breve{\ba}, \Sigma_{\hat{\ba}} + \Sigma_{\breve{\ba}})}{\norm(\hat{\ba} \given \hat{\ba}, \Sigma_{\hat{\ba}})} \quad .
\end{eqnarray}
We do not explore non-Gaussian priors, as these are considerably more difficult to implement.

\subsection{Practical usage}
The \texttt{Cloudy} implementation in \fsps{} includes 128 emission lines; any or all of them may be optimized or marginalized over using this scheme. In practice, it is suggested to only optimize or marginalize over emission lines included in an observed spectrum. The marginalization calculation is fast and adds little computational overhead in most cases. Note that neither optimization nor marginalization respects emission line physics. This means that, for example, emission line doublets may be assigned forbidden ratios if the data prefer it. The added emission lines may also have {\it negative} luminosities. In objects with weak or no nebular emission, the user may wish to turn off the optimization or limit it to strong emission features to prevent it from instead erroneously modeling differences between the model and observed continuum. This is particularly true in cases where a weak or nonexistent emission line lies on top of an informative absorption feature, such as H$\alpha$ or H$\beta$.

When applying the Gaussian priors, the center of the prior for each line is taken as the prediction from the \texttt{Cloudy} implementation. This means that the lines are assumed to be powered by normal star formation, but deviations are permitted. The width of the prior is specified by the user, in units of (true/predicted) luminosity.

\end{appendix}


\end{document}